\documentclass[a4paper,11pt]{article} 
\pdfoutput=1
\usepackage{jheppub}
\usepackage{amsmath,amssymb}
\usepackage{hhline}
\usepackage{subfigure}
\usepackage{epstopdf}

\newcommand{\ba}{\begin{eqnarray}}
\newcommand{\ea}{\end{eqnarray}}
\newcommand{\be}{\begin{equation}}
\newcommand{\ee}{\end{equation}}
\newcommand{\bal}{\begin{align}}
\newcommand{\eal}{\end{align}}

\title{\boldmath 
$B_{s}\to K \ell \nu_\ell$ and $B_{(s)} \to \pi (K) \ell^+\ell^-$
decays \\[2mm]  at large 
recoil and CKM matrix elements }

\author[a]{Alexander~Khodjamirian} 
\author[a,b]{and Aleksey V. Rusov} 
\affiliation[a]{Theoretische Physik 1,
  Naturwissenschaftlich-Technische Fakult\"at, Universit\"at Siegen, D-57068 Siegen, Germany}
\affiliation[b]{
Department of Theoretical Physics,
P.G. Demidov Yaroslavl State University, 150000, Yaroslavl, Russia
}
\emailAdd{khodjamirian@physik.uni-siegen.de}
\emailAdd{rusov@physik.uni-siegen.de}

\preprint{SI-HEP-2017-03, QFET-2017-03 }

\abstract{ 
We provide hadronic input  for the $B$-meson 
semileptonic transitions to a light pseudoscalar meson 
at large recoil. The $B_s\to K$ form factor 
calculated from QCD light-cone sum rule is updated, 
to be used for a  $|V_{ub}|$
determination from the $B_s\to K \ell \nu$ width. 
Furthermore,  we calculate the hadronic input for  the binned 
observables of $B \to \pi \ell^+ \ell^-$ 
and $B \to K \ell^+ \ell^-$. In addition to the form factors, the nonlocal 
hadronic matrix elements are obtained, combining 
QCD factorization and light-cone sum rules with hadronic dispersion relations.
We emphasize that, due to nonlocal effects, the ratio of branching 
fractions  of these decays is not sufficient for an accurate extraction of  
the $|V_{td}/V_{ts}|$ ratio.  Instead, we suggest to determine   
the Wolfenstein parameters $A,\rho,\eta$ of the CKM matrix, 
combining the branching fractions of $B \to K \ell^+ \ell^-$ 
and $B \to \pi \ell^+ \ell^-$ with the direct $CP$-asymmetry 
in the latter decay. We also obtain the hadronic matrix elements for a yet unexplored channel 
$B_s \to K \ell^+ \ell^-$.
}


\begin{document}

\maketitle
\flushbottom

\section{Introduction}

Determination of CKM matrix elements from the semileptonic decays of $B$ meson  
remains a topical problem. Most importantly, 
one has to clarify the origin of the tension between 
the $|V_{ub}|$ values extracted from the exclusive $B\to \pi\ell\nu_\ell$ 
and inclusive $B\to X_u \ell \nu_\ell$ decays (see e.g., the review \cite{pdgVubVcb}).
The $B\to \pi$ vector form factor  $f^+_{B\pi}(q^2)$   
is the only  theory input sufficient for 
the $|V_{ub}|$ determination from   $B\to \pi\ell\nu_\ell$. 
This hadronic matrix element is calculated in the lattice QCD 
at small recoil of the pion (at large $q^2$) 
or from  QCD light-cone sum rules 
(LCSRs) at large  recoil of the pion (at small and intermediate $q^2$).

Apart from increasing the accuracy of the form factor calculation,
it is important to extend the set of ``standard'' exclusive 
processes used for $|V_{ub}|$ determination.
The $B_s \to  K^* (\to K \pi) \ell \nu_\ell$ decay, as one possibility,
was discussed in \cite{Feldmann:2015xsa}.
A simpler process is the  $B_s\to K\ell\nu_\ell$ decay, where the data 
are anticipated from LHCb collaboration.  
Our first goal  in this paper is to provide 
this decay  mode with a hadronic input, updating the
calculation of the $B_s\to K$ form factors from LCSRs.
This method \cite{Balitsky:1986st,Balitsky:1989ry,Chernyak:1990ag} is
based on the operator-product expansion (OPE)
of a correlation function expressed in terms of light-meson
distribution amplitudes (DAs) with growing twist. 
The violation of the $SU(3)_{fl}$ symmetry 
in  $B_s\to K$  with respect to $B\to \pi$ transition  
emerges in LCSRs due to the $s$-quark mass effects in 
the correlation function, including  the asymmetry between the $s$- and $\{u,d\}$-partons 
in the kaon DAs. 
Earlier LCSR results on the $B_s\to K$  form factors 
can be found in \cite{Duplancic:2008tk}, where the NLO corrections 
to the correlation function computed in 
\cite{Duplancic:2008ix} were taken into account.
In this paper we update the LCSRs
for   $f^+_{B_s K}(q^2)$ and also for the tensor form factor 
 $f^T_{B_s K}(q^2)$. In particular, we correct 
certain terms in the subleading twist-3,4 contributions to LCSRs for both vector
and tensor form factors. In parallel, 
we recalculate the $B\to K$ and $B\to \pi$ form factors 
using a common set of input parameters, e.g.,  
the updated \cite{Gelhausen:2013wia} 2-point QCD sum rule for the decay constants 
$f_B$ and $f_{B_s}$.  Importantly, the 
twist-5,6 corrections to the LCSRs estimated by one of us \cite{Rusov_tw56}  
are negligibly small, ensuring the reliability 
of the adopted twist $\leq 4$ approximation.

The calculated form factors are then used to address 
the second goal of this paper: determination of CKM parameters 
from  the flavour-changing neutral current (FCNC) 
decays $B \to K \ell^+ \ell^-$, $B \to \pi \ell^+ \ell^-$ and 
$B_s \to K \ell^+ \ell^-$.
Recently, $|V_{td}|$, $|V_{ts}|$ and their ratio were determined 
by LHCb collaboration \cite{LHCb:Bpimumu15}  
from the measured $B \to \pi \ell^+ \ell^-$ 
and $B \to K \ell^+ \ell^-$ partial widths. 
We suggest to make the extraction of CKM parameters
from these decays more accurate and comprehensive.    
As well known, in addition to the semileptonic form factors,  
the hadronic input in FCNC decays 
includes  also nonlocal hadronic matrix elements
emerging due to the electromagnetic lepton-pair emission
combined with the weak transitions. These hadronic matrix elements 
in the $B \to \pi \ell^+ \ell^-$ decay amplitude
are multiplied by the CKM parameters other than $V_{td}$,  
making the determination of the latter not straightforward. 
We take into account
the nonlocal hadronic effects in  $B \to K \ell^+ \ell^-$ and  
$B \to \pi \ell^+ \ell^-$, employing the methods used 
in \cite{Khodjamirian:2012rm}, \cite{Hambrock:2015wka} and 
originally suggested in \cite{Khodjamirian:2010vf}. 
The nonlocal hadronic matrix elements 
are calculated at spacelike $q^2$,   
using OPE, QCD factorization \cite{Beneke:2001at} and LCSRs,  
and are then matched to their values 
at timelike $q^2$  via hadronic dispersion relations.
The results of this calculation are reliable at large hadronic recoil, 
below the charmonium region, that is, at $q^2<m_{J/\psi}^2$.
Here we also extend the calculation of nonlocal effects to the 
previously unexplored channel $B_s \to K \ell^+ \ell^-$.  

The binned widths and direct $CP$-asymmetries of FCNC semileptonic decays
are then expressed  in a form combining the CKM parameters 
with  the quantities determined by the  calculated 
hadronic input. Here we find it more convenient to switch 
to the Wolfenstein parametrization of the CKM matrix. In this form, three observables:
the width of $B \to K \ell^+ \ell^-$,  the ratio of 
$B \to \pi \ell^+ \ell^-$ and $B \to K \ell^+ \ell^-$ widths 
and the direct $CP$-asymmetry in $B \to \pi \ell^+ \ell^-$, are sufficient 
to extract the three Wolfenstein parameters $A$, $\eta$ and $\rho$ 
from experimental data, provided the parameter $\lambda$ 
is known quite precisely. Two additional observables for the same
determination are given by the yet unobserved $B_s \to K\ell^+\ell^-$ decay.  
The current data on the $B \to K \ell^+ \ell^-$ and
$B \to \pi \ell^+ \ell^-$ decays are not yet precise enough to yield the CKM parameters
with an accuracy comparable to the other determinations. 
Hence, here we limit ourselves with the Wolfenstein parameters taken 
from the global CKM fit and predict the binned observables of all three
FCNC decays  in the optimal interval 
$1.0 \mbox{ GeV}^2 < q^2 < 6.0 \mbox{ GeV}^2$ of the large recoil region.

In what follows, in Sect.~2 we specify and discuss 
the hadronic input and observables in the exclusive semileptonic 
$B_{(s)}$ decays. In Sect.~3 we present the numerical results and Sect.~4 is devoted to 
the final discussion. In the Appendices,   
we briefly recapitulate the  calculation  (A) of the  
form factors from LCSRs and (B) of the nonlocal hadronic matrix elements.  

\section{Observables in semileptonic {\boldmath $B_{(s)}$} decays and CKM parameters}

The form factors of semileptonic transitions 
of $B$-meson to a light pseudoscalar meson $P = \pi, K$
are defined in a standard way:

\begin{equation}
\label{eq:fpl}
\! \langle P (p) | \bar q \gamma^\mu b | B (p+q) \rangle =  
f^+_{B P}  (q^2) \left [2 p^\mu + \left(1 - \frac{m_B^2 - m_P^2}{q^2} \right) \! q^\mu \right] 
+ f^0_{B P}  (q^2) \, \frac{m_B^2 - m_P^2}{q^2} \, q^\mu,  
\end{equation}
\begin{equation}
\label{eq:fT}
\langle P (p) | \bar q \sigma^{\mu\nu} q_\nu b | B (p + q) \rangle =  
\frac{i f^T_{B P} (q^2)} {m_B + m_P} 
\left[ 2q^2 p^\mu + \bigg(q^2 - \left ( m_B^2 - m_P^2 \right )\bigg) q^\mu \right], 
\end{equation}
where $p^\mu$ and $q^\mu$ are the four-momenta of the 
$P$-meson
and lepton pair, respectively, and the vector and scalar form factors
coincide at $q^2=0$, that is, $f^+_{B P}  (0)=f^0_{B P}  (0)$. 

We start from the weak semileptonic decay $\bar B_s\to K^+ \ell \bar{\nu}_\ell$,   
where the hadronic input for $\ell=e,\mu$ in the  $m_\ell=0$
approximation is given by the vector form factor $f^+_{B_s K}$. 
We use the following quantity related to the differential width 
integrated over an interval $0\leq q^2\leq q_0^2$:
\begin{equation}
\!\!\Delta\zeta_{B_s K}\,[0,q_{0}^2]\equiv
\frac{G_F^2}{24\pi^3}\int\limits_0^{q_{0}^2}dq^2p_{B_sK}^3
|f_{B_s K}^+(q^2)|^2= \frac{1}{|V_{ub}|^2\tau_{B_s}}
\int\limits_0^{q_{0}^2} dq^2\frac{d B (\bar{B}_s\to K^+ \ell \bar{\nu}_\ell)}{dq^2}\,,
\label{eq:zeta} 
\end{equation}
where the $q^2$-dependent kinematical factor 
$p_{BP} = [(m_B^2+m_P^2-q^2)^2/(4m_B^2)- m_P^2]^{1/2}$ is the 3-momentum of 
$P$ meson in the rest frame of $B$ meson. Our choice for the
integration interval is  $q_0^2=12.0~\mbox{GeV}^2$, 
covering the region where the LCSRs used for the calculation of the form factors
(see Appendix A)  are valid. The same  
interval was adopted for  the 
analogous quantity $\Delta\zeta_{B \pi}[0,q_0^2]$  for $B\to \pi\ell\nu_\ell$  
calculated in \cite{Khodjamirian:2011ub,Imsong:2014oqa}. 
The numerical estimate of $\Delta\zeta_{B_s K}\,[0,q_0^2]$ 
presented in the next section can be directly used for $|V_{ub}|$ determination, 
provided the integrated branching fraction on the r.h.s. of Eq.~(\ref{eq:zeta}) is  measured.

Turning to semileptonic decays generated by the $b\to s (d) \ell^+\ell^-$
transitions ($\ell=e,\mu$), we use a generic notation  $\bar{B} \to P \ell^+ \ell^-$ for the 
three channels: $B^-\to K^- \ell^+\ell^-$,  
$B^-\to \pi^- \ell^+\ell^-$ 
and  $\bar{B}_s\to K^0 \ell^+\ell^-$
\footnote{For simplicity we 
consider a transition into the fixed flavour state $K^0$ which is easy 
to convert to $K_s$ if needed.}
, denoting the $CP$ conjugated 
channels by $B \to \bar{P} \ell^+ \ell^-$. 
The decay amplitude can be represented in the following form:
\begin{eqnarray}
\hspace*{-3mm} A(\bar{B} \to P \ell^+ \ell^-) & = &  
\frac{G_F}{\sqrt 2} \frac{\alpha_{\rm em}}{\pi} 
\Bigg\{ \Big[\lambda_t^{(q)} f^+_{B P} (q^2)c_{BP}(q^2)
+ \lambda_u^{(q)}d_{B P} (q^2)\Big]
\left(\bar \ell \gamma^\mu \ell \right)  p_\mu 
\nonumber \\
& + & \lambda_t^{(q)} C_{10} f^+_{B P} (q^2)\left(\bar \ell \gamma^\mu \gamma_5 \ell \right) 
p_\mu \Bigg\}\,,
\label{eq:ampl}
\end{eqnarray}
where $ \lambda_p^{(q)}= V_{pb}V^*_{pq}$  ($p=u,c,t$; $q=d,s$), $m_\ell=0$, 
and we use unitarity of the CKM matrix, fixing hereafter 
$\lambda_c^{(q)}=-(\lambda_t^{(q)}+ \lambda_u^{(q)})$.
In Eq.~(\ref{eq:ampl}) we introduce 
a compact notation:
\begin{eqnarray}
c_{BP}(q^2)= C_9+ \frac{2 (m_b + m_q)}{m_B + m_P} C_7^{\rm eff}
\frac{f^T_{B P} (q^2)}{f^+_{B P} (q^2)} +
16\pi^2\frac{{\cal H}_{BP}^{(c)}(q^2)}{f^+_{B P} (q^2)}\,,
\label{eq:amplc}
\end{eqnarray}
where $m_q$ is the mass of $d$ or $s$-quark 
and 
\begin{eqnarray}
d_{BP}(q^2)= 16\pi^2\Big({\cal H}_{BP}^{(c)}(q^2)-{\cal H}_{BP}^{(u)}(q^2)\Big)\,.
\label{eq:amplh}
\end{eqnarray}
In addition, we introduce the phase difference 
of the hadronic  amplitudes  defined above: 
\begin{equation}
\delta_{BP}(q^2)= {\rm Arg} (d_{BP}(q^2)) - {\rm Arg} (c_{BP} (q^2)). 
\end{equation}
In Eq.~(\ref{eq:ampl}) the dominant contributions of the  
operators $O_{9,10}$ and $O_{7\gamma}$  of the effective Hamiltonian (see Appendix B) 
are expressed in terms of the vector and tensor $B\to P$ form factors,
$f^+_{B P} (q^2)$ and $f^T_{B P} (q^2)$, respectively, 
defined in Eqs.~(\ref{eq:fpl}) and (\ref{eq:fT}). 
The amplitudes ${\cal H}_{BP}^{(c,u)}(q^2)$ 
parametrize the nonlocal contributions to $B\to P \ell^+\ell^-$, 
generated by the  current-current, quark-penguin 
and chromomagnetic operators  in the effective Hamiltonian,
combined with an electromagnetically produced  lepton pair.
The definition of nonlocal amplitudes is given in Appendix B,
where also the method of their calculation
is briefly explained. 
In Refs.~\cite{Khodjamirian:2010vf,Khodjamirian:2012rm},
this part of hadronic input was cast 
in the form  of an effective (process- and $q^2$-dependent)
addition $\Delta C_9^{BP}(q^2) $ to the Wilson coefficient $C_9$.
Here, as in Ref.~\cite{Hambrock:2015wka}, we find it more convenient 
to separate the parts proportional to $\lambda_u^{(q)}$ and 
$\lambda_c^{(q)} =  - (\lambda_t^{(q)}+\lambda_u^{(q)})$. 

Squaring  the amplitude (\ref{eq:ampl}) and integrating over the
phase space, one obtains for the $q^2$-binned branching fraction, defined as:
\begin{equation}
{\cal B}( \bar B \to P\ell^+\ell^-[q_1^2,q_2^2])\equiv
\frac{1}{q_2^2-q_1^2}\int\limits_{q_1^2}^{q_2^2}dq^2
\frac{d B  (\bar{B} \to P\ell^+ \ell^-)}{d q^2}\,,
\label{eq:bin}
\end{equation}
the following expression:
\begin{eqnarray}
\label{eq:Brbin}
{\cal B}(\bar{B}\to P\ell^+\ell^-[q_1^2,q_2^2])=
\frac{G_F^2 \alpha_{\rm em}^2 |\lambda_t^{(q)}|^2}{192\pi^5} 
\Bigg\{ {\cal F}_{BP}[q_1^2,q_2^2]+ \kappa_q^2{\cal D}_{BP}[q_1^2,q_2^2]
\nonumber\\
+ 2 \kappa_q\Big(\cos \xi_q \, {\cal C}_{BP}[q_1^2,q_2^2] 
- \sin \xi_q \, {\cal S}_{BP}[q_1^2,q_2^2]
\Big)
\Bigg\}\tau_{B}\,,
\end{eqnarray}
where the ratio of CKM matrix elements is parametrized
in terms of its module and phase: 
\begin{equation}
\label{eq:kappa}
\frac{\lambda_u^{(q)}}{\lambda_t^{(q)}} = 
\frac{V_{ub} V_{uq}^*}{V_{tb} V_{tq}^*} \equiv \kappa_q \, e^{i \xi_q}, ~~(q=d,s)\,,
\end{equation} 
and we use  the following notation for the phase-space weighted 
and integrated  parts of the decay amplitude squared:
\begin{equation}
{\cal F}_{BP}[q_1^2,q_2^2] = \frac{1}{q_2^2 - q_1^2}
\int\limits_{q_1^2}^{q_2^2} d q^2 \, p_{BP}^3
|f^+_{B P}(q^2)|^2 \Big(\left|c_{BP}(q^2)\right|^2 + |C_{10}|^2 \Big)\,,
\label{eq:binF}
\end{equation}
\begin{equation}
{\cal D}_{BP}[q_1^2,q_2^2] = \frac{1}{q_2^2 - q_1^2}
\int\limits_{q_1^2}^{q_2^2} d q^2 \, p_{BP}^3
\left|d_{BP}(q^2)\right|^2\,,
\label{eq:binD}
\end{equation}
\begin{equation}
{\cal C}_{BP}[q_1^2,q_2^2] = \frac{1}{q_2^2 - q_1^2}
\int\limits_{q_1^2}^{q_2^2} d q^2 \, p_{BP}^3
\left|f_{BP}^+ (q^2) c_{BP}(q^2)d_{BP}(q^2)\right|\cos\delta_{BP}(q^2)\,,
\label{eq:binC}
\end{equation}
\begin{equation}
{\cal S}_{BP}[q_1^2,q_2^2] = \frac{1}{q_2^2 - q_1^2}
\int\limits_{q_1^2}^{q_2^2} d q^2 \, p_{BP}^3
\left|f_{BP}^+ (q^2) c_{BP}(q^2)d_{BP}(q^2)\right|\sin \delta_{BP}(q^2)\,.
\label{eq:binS}
\end{equation}
The binned branching fraction  for the $CP$-conjugated mode 
$B\to \bar{P} \ell^+\ell^-$ is obtained from Eq.~(\ref{eq:Brbin}) 
by changing the sign at the  term proportional to $\sin \xi_q$.

Furthermore, we 
consider  two binned observables: the $CP$-averaged branching fraction:
\begin{eqnarray}
\label{eq:Brbinav}
{\cal B}_{BP}[q_1^2,q_2^2]\equiv \frac{1}{2}\Big({\cal B}(\bar{B}\to P\ell^+\ell^-[q_1^2,q_2^2])
+ {\cal B}(B\to \bar{P}\ell^+\ell^-[q_1^2,q_2^2])\Big)
\nonumber\\
=
\frac{G_F^2 \alpha_{\rm em}^2 |\lambda_t^{(q)}|^2}{192\pi^5} 
\Bigg\{ {\cal F}_{BP}[q_1^2,q_2^2] + \kappa_q^2 \, {\cal D}_{BP}[q_1^2,q_2^2]
+ 2 \kappa_q \cos\xi_q \, {\cal C}_{BP}[q_1^2,q_2^2]
\Bigg\}\tau_{B},
\end{eqnarray}
and the corresponding direct $CP$-asymmetry:
\begin{eqnarray}
\label{eq:Acpbin}
{\cal A}_{BP}[q_1^2,q_2^2]= \frac{
{\cal B}(\bar{B}\to P\ell^+\ell^-[q_1^2,q_2^2])
- 
{\cal B}(B\to  \bar{P}\ell^+\ell^-[q_1^2,q_2^2])}{
{\cal B}(\bar{B}\to P\ell^+\ell^-[q_1^2,q_2^2])
+{\cal B}(B\to \bar{P}\ell^+\ell^-[q_1^2,q_2^2])}
\nonumber\\
=
\frac{- 2 \kappa_q \sin\xi_q\, {\cal S}_{BP}[q_1^2,q_2^2]}{
{\cal F}_{BP}[q_1^2,q_2^2]+ \kappa^2_q \, {\cal D}_{BP}[q_1^2,q_2^2]
+ 2 \kappa_q \cos \xi_q\, {\cal C}_{BP}[q_1^2,q_2^2] }\,.
\label{eq:cpasymm}
\end{eqnarray}
In Eqs.~(\ref{eq:Brbinav}) and (\ref{eq:cpasymm}) 
the CKM-dependent coefficients are conveniently separated 
from the quantities 
${\cal F}_{BP}$, ${\cal D}_{BP}$, ${\cal C}_{BP}$, ${\cal S}_{BP}$, 
which contain the calculable hadronic matrix elements, Wilson coefficients 
and kinematical factors. In the next section we  present numerical results 
for these quantities  for a definite $q^2$-bin in the 
large-recoil region.

Turning to the observables for the specific decay channels, we neglect 
$\lambda_u^{(s)}$, hence, put $\kappa_s=0$ and obtain for $B\to K \ell^+\ell^-$:
\begin{eqnarray}
\label{eq:BrBK}
{\cal B}_{BK}[q_1^2,q_2^2]=
\frac{G_F^2 \alpha_{\rm em}^2 |\lambda_t^{(s)}|^2}{192\pi^5} 
{\cal F}_{BK}[q_1^2,q_2^2]\tau_{B},
\end{eqnarray}
with vanishing $CP$ asymmetry.
For $B^- \to \pi^- \ell^+\ell^-$ and its $CP$-conjugated process 
both observables,
\begin{eqnarray}
\label{eq:BrBpi}
{\cal B}_{B\pi}[q_1^2,q_2^2]=
\frac{G_F^2 \alpha_{\rm em}^2 |\lambda_t^{(d)}|^2}{192\pi^5} 
\Bigg\{ {\cal F}_{B\pi}[q_1^2,q_2^2]
+ \kappa_d^2 \, {\cal D}_{B\pi}[q_1^2,q_2^2]
+ 2 \kappa_d \cos \xi_d\,{\cal C}_{B\pi}[q_1^2,q_2^2] 
\Bigg\}\tau_B\,,
\nonumber
\\
\end{eqnarray}
and 
\begin{eqnarray}
\label{eq:AcpBpi}
{\cal A}_{B\pi}[q_1^2,q_2^2]=\frac{- 2 \kappa_d\,\sin\xi_d \, {\cal S}_{B\pi}[q_1^2,q_2^2]
}{
{\cal F}_{B\pi}[q_1^2,q_2^2]+ \kappa^2_d \, {\cal D}_{B\pi}[q_1^2,q_2^2]
+ 2 \kappa_d \,\cos \xi_d \, {\cal C}_{B\pi}[q_1^2,q_2^2] }\,,
\end{eqnarray}
are relevant. 
The corresponding observables 
${\cal B}_{B_sK}[q_1^2,q_2^2]$ and ${\cal A}_{B_s K}[q_1^2,q_2^2]$
for $\bar{B}_s\to K^0 \ell^+\ell^-$ and its $CP$-conjugated mode
are given by the expressions similar to Eqs.~(\ref{eq:BrBpi}), (\ref{eq:AcpBpi}), 
with  $B\pi$ replaced by $B_sK$. Here we do not consider
the decays $\bar{B}^0\to \bar{K}^0 \ell^+\ell^-$  and  
$\bar{B}^0\to \pi^0 \ell^+\ell^-$, which are the isospin
counterparts of, respectively, 
$B^-\to K^- \ell^+\ell^-$ and  
$B^- \to \pi^- \ell^+\ell^-$ and can be treated in a 
similar way (see \cite{Khodjamirian:2012rm,Hambrock:2015wka}).
We also postpone to a future study the time-dependent $CP$-asymmetry
in the $\bar{B}_s\to K^0\ell^+\ell^-$ decay.

Dividing  Eq.~(\ref{eq:BrBpi}) by Eq.~(\ref{eq:BrBK}), 
we notice that an
accurate extraction of the ratio $ |V_{td}/V_{ts}|$ from the ratio 
of branching fractions ${\cal B}_{B\pi}[q_1^2,q_2^2]/{\cal B}_{B\pi}[q_1^2,q_2^2]$
can only be achieved if the contributions 
of process-dependent nonlocal effects  are taken into account for both decay
modes. Moreover, this ratio depends also on the other CKM
parameters, most importantly, on the $V_{ub}$ value
\footnote{Note that in the analysis of $B \to \pi \ell^+\ell^-$  and 
$B\to K \ell^+\ell^-$ presented in \cite{LHCb:Bpimumu15} 
these effects are not explicitly specified.}.

Here we suggest a different, more systematic way to 
extract the parameters of CKM matrix from the observables 
(\ref{eq:BrBK})-(\ref{eq:AcpBpi}).  First of all, we find it 
more convenient to switch to the four standard 
Wolfenstein parameters $\lambda$, $A$, $\rho$ and $\eta $
defined as in \cite{pdg}. 
The  relevant CKM factors can be represented as follows:
\begin{eqnarray}
\lambda_{t}^{(s)} = - A \lambda^2\,, 
\label{eq:lamts}\\
\left|\frac{\lambda_{t}^{(d)}}{\lambda_{t}^{(s)}} \right| =
\left|\frac{V_{td}}{V_{ts}}\right| = \lambda \sqrt{(1-\rho)^2 +\eta^2},
\label{eq:lamtd}
\end{eqnarray}
\begin{equation}
\frac{\lambda_{u}^{(d)}}{\lambda_{t}^{(d)}} = \frac{V_{ub} V_{ud}^*}{V_{tb} V_{td}^*}  
\equiv \kappa_d e^{i \xi_d} 
= \left(1 - \frac{\lambda^2}{2} \right) 
\frac{\rho (1 - \rho) - \eta^2 - i \eta}{(1-\rho)^2 + \eta^2},
\label{eq:kapdef}
\end{equation}
so that
\begin{eqnarray}
\kappa_d  =  \left(1 - \frac{\lambda^2}{2} \right) 
\frac{\sqrt{(\rho (1 - \rho) - \eta^2)^2 + \eta^2}}{(1-\rho)^2 +
  \eta^2}, 
\label{eq:kapdW}
\\
\sin \xi_d  =  
\frac{-\eta}{\sqrt{(\rho(1-\rho)-\eta^2)^2 + \eta^2}},~~
\cos \xi_d  =  
\frac{\rho(1-\rho)-\eta^2}{\sqrt{(\rho(1-\rho)-\eta^2)^2 + \eta^2}},
\label{eq:sincosW}
\end{eqnarray}
where we neglect very small $O(\lambda^4)$ corrections to these 
expressions
\footnote{This is consisent with neglecting the 
$O(\lambda_u^{(s)})\sim O(\lambda^4)$ terms
 in the $B\to K \ell^+\ell^-$  amplitude. These terms contain
 nonlocal effects generated by the $u$-quark loops and calculable  
within our approach. Hence, achieving the $O(\lambda^4)$ precision 
is possible in future.}. 

Hereafter, we suppose, that the parameter  $\lambda$, 
precisely determined from the global CKM fit \cite{pdg},
is used as an input. Then, it is possible to extract  all three 
remaining Wolfenstein parameters combining the three observables 
(\ref{eq:BrBK})-(\ref{eq:AcpBpi})  for semileptonic FCNC decays. 
First, the parameter  $A$ is determined from the 
binned branching fraction of $B\to K \ell^+\ell^-$, as follows 
after substituting  Eq.~(\ref{eq:lamts}) in Eq.~(\ref{eq:BrBK}):
\begin{equation}
A=
\frac{(192\pi^5) ^{1/2}}{G_F\alpha_{em}\lambda^2}
\Bigg(\frac{1}{{\cal F}_{BK}[q_1^2,q_2^2]}\Bigg)^{1/2}
\Bigg(\frac{{\cal B}_{BK}[q_1^2, q_2^2]}{\tau_B}\Bigg)^{1/2}\!.
\label{eq: detA}
\end{equation}
Then, combining the ratio of the $B\to \pi \ell^+\ell^-$ and
$B\to K \ell^+\ell^-$ binned branching fractions with the $CP$-asymmetry
of the pion mode, and employing Eqs.~(\ref{eq:lamtd}), (\ref{eq:kapdW}) and (\ref{eq:sincosW}), 
we obtain for the parameter $\eta$ the following relation: 
\begin{equation}
\eta=\frac{1}{2\lambda^2(1-\lambda^2/2)}
 \Bigg(\frac{{\cal F}_{BK}[q_1^2,q_2^2]}{{\cal S}_{B\pi }[q_1^2,q_2^2]}\Bigg)
\Bigg({\cal A}_{B\pi}[q_1^2,q_2^2]
\frac{{\cal B}_{B\pi}[q_1^2,q_2^2]}{{\cal B}_{BK}[q_1^2,q_2^2]}\Bigg)\,.
\label{eq:etadet}
\end{equation}
Finally, after $\eta$ is determined, the parameter $\rho$ can be 
extracted from the ratio of branching fractions 
(\ref{eq:BrBpi}) 
and (\ref{eq:BrBK})  written explicitly in terms of $\eta$ and $\rho$:  
\begin{eqnarray}
\frac{{\cal B}_{B\pi}[q_1^2,q_2^2]}{{\cal B}_{BK}[q_1^2,q_2^2]}=
\frac{\lambda^2}{{\cal F}_{BK}[q_1^2,q_2^2]} 
\Bigg(\left[(1-\rho)^2+\eta^2\right] {\cal F}_{B\pi} [q_1^2,q_2^2]
\nonumber\\
+
\frac{ \left[\rho(1-\rho)-\eta^2\right]^2 +\eta^2}{(1-\rho)^2+\eta^2} \left(1-\frac{\lambda^2}{2}\right)^2{\cal
  D}_{B\pi}[q_1^2,q_2^2]
\nonumber\\
+
2\left[ \rho(1-\rho)-\eta^2 \right] \left(1-\frac{\lambda^2}{2}\right){\cal C}_{B\pi}[q_1^2,q_2^2]
\Bigg)\,.
\label{eq:rhodet}
\end{eqnarray}
Similar  relations for the $B_s\to K \ell^+\ell^-$ decay,
obtained by replacing $B\pi\to B_sK$ in Eqs.~(\ref{eq:etadet}) 
and (\ref{eq:rhodet}), provide an additional source of these parameters. 

\section{Numerical Results}
\begin{table}[t]
\begin{center}
\begin{tabular}{|c|c|}
\hline
Parameter & Ref. \\
\hline
$G_F= 1.1664\times 10^{-5}~\mbox{GeV}^2$; ~~$\alpha_{em}=1/129$ & \\
$\alpha_{s}(m_Z)=0.1185\pm 0.0006$; $\alpha_{s}(3~\mbox{GeV} )=  0.252$ & \cite{pdg} \\
$\overline{m}_b(\overline{m}_b)=4.18\pm 0.03$ GeV;~~$\overline{m}_c(\overline{m}_c)=1.275\pm 0.025$ GeV &\\
$\overline{m}_s(2 ~\mbox{GeV})= 95 \pm 10$  MeV & \\
\hline
$\mu=3.0_{-0.5}^{+1.5}$ GeV& \\
\hline
$f_{\pi}=130.4$ MeV; ~~$f_K=159.8$ MeV & \cite{pdg} \\
$a_{2}^{\pi}(1 \mbox{GeV}) =0.17 \pm 0.08$; ~$a_{4}^{\pi}(1 \mbox{GeV})= 0.06\pm 0.10$ &\cite{Khodjamirian:2011ub}\\
$a_1^{K}(1 \mbox{GeV})=0.10\pm 0.04$;~ $a_2^{K}(1 \mbox{GeV})=0.25\pm 0.15$    &\cite{Chetyrkin:2007vm, Khodjamirian:2009ys} \\
 $\mu_\pi(2~ \mbox{GeV})= 2.50 \pm 0.30$ GeV; ~$\mu_K(2~\mbox{GeV})= 2.49 \pm 0.26$ GeV  & 
\cite{Leutwyler:1996qg, Khodjamirian:2009ys}\\
$M^2=16 \pm 4 $ GeV$^2$ ($M^2=17 \pm 4 $ GeV$^2$) [in $B$($B_s$)-channel] &
\cite{Khodjamirian:2011ub} \\ 
\hline
$\lambda_B=460 \pm 110$ MeV  & \cite{Braun:2003wx} \\
$M^2=1.0\pm 0.5 $ GeV$^{2}$;~ 
$s_0^\pi=0.7 $ GeV$^{2}$;~ 
$s_0^K= 1.05$ GeV$^{2}$  
& \cite{Khodjamirian:2010vf} \\
\hline
\end{tabular}
\caption{Input parameters used in the numerical analysis.}
\label{tab:input}
\end{center}
\end{table}
The most important input parameters used in our numerical analysis 
are listed in Table~\ref{tab:input}. In particular, the electroweak parameters,
the strong coupling and the meson masses are taken from \cite{pdg}. 
For the quark masses in $\overline{MS}$ scheme, entering the correlation functions  
for QCD sum rules, we adopt, following, e.g., \cite{Gelhausen:2013wia}, the 
intervals covering the non-lattice determinations in \cite{pdg}. 
We put $m_{u,d} = 0$, except in the combination $\mu_{\pi(K)}= m_{\pi(K)}^2/(m_u+m_{d(s)})$
entering the pion and kaon DAs.
In LCSRs, parameters of the pion and kaon twist-2 DA's include 
the decay constants, and the  Gegenbauer moments $a^\pi_{2,4}$ and $a^K_{1,2}$.
Normalization  of the twist-3 DAs is determined by $\mu_{\pi,K}$,
where the  ChPT relations \cite{Leutwyler:1996qg} between light-quark masses 
are used (see e.g., \cite{Khodjamirian:2009ys}).
The remaining  parameters of the twist-3 and twist-4 DAs, 
not shown  in Table \ref{tab:input} for brevity, are taken 
from \cite{Ball:2006}, they were also used in 
\cite{Khodjamirian:2011ub,Khodjamirian:2010vf,Khodjamirian:2009ys}.
Furthermore, in LCSRs  the renormalization scale $\mu$ 
and the Borel parameters $M$ for the sum rules 
with $B$ ($B_s$) interpolating current quoted in Table~\ref{tab:input} 
are chosen, largely following \cite{Khodjamirian:2011ub}.
The effective quark-hadron duality threshold is determined
calculating the $B_{(s)}$-meson mass 
from the differentiated LCSR.  The decay constants $f_B$ and $f_{B_s}$
entering LCRSs are replaced by the two-point sum rules in NLO, their 
expressions and input parameters (in particular, the vacuum condensate 
densities) are the same as in \cite{Gelhausen:2013wia}.
The intervals obtained from these sum 
rules in NLO are 
$f_B = (202^{+35}_{-21}) \, {\rm MeV}$, 
$f_{B_s} = (222^{+38}_{-24})\, {\rm MeV}$.  
Note that the above uncertainties are effectively smaller in 
LCSRs (Eq.~(\ref{eq:lcsrs}) in Appendix A) due to the correlations of common
parameters.
    
\begin{figure}[h]\center
\includegraphics[scale=0.5]{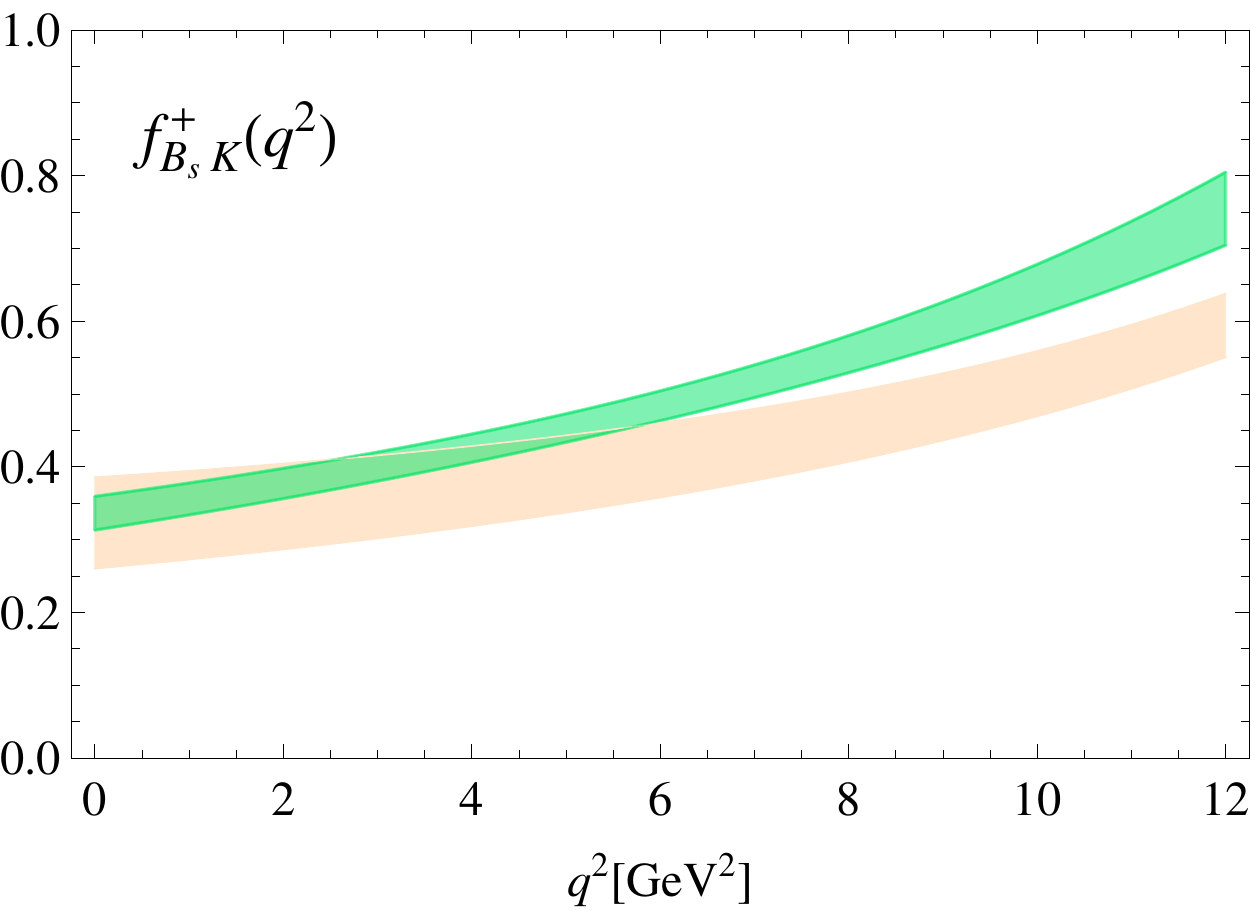} \quad
\includegraphics[scale=0.5]{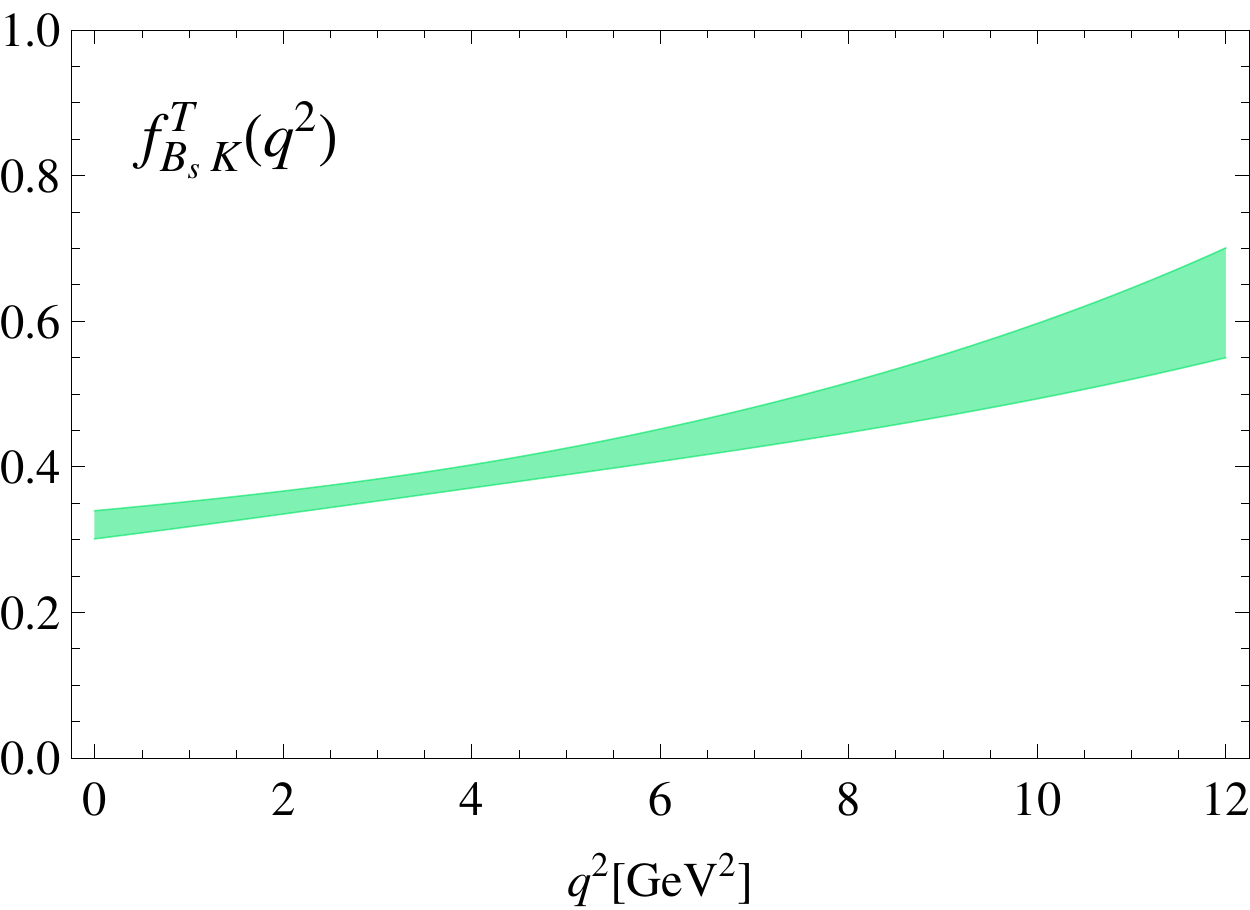} \\[2mm]
\includegraphics[scale=0.5]{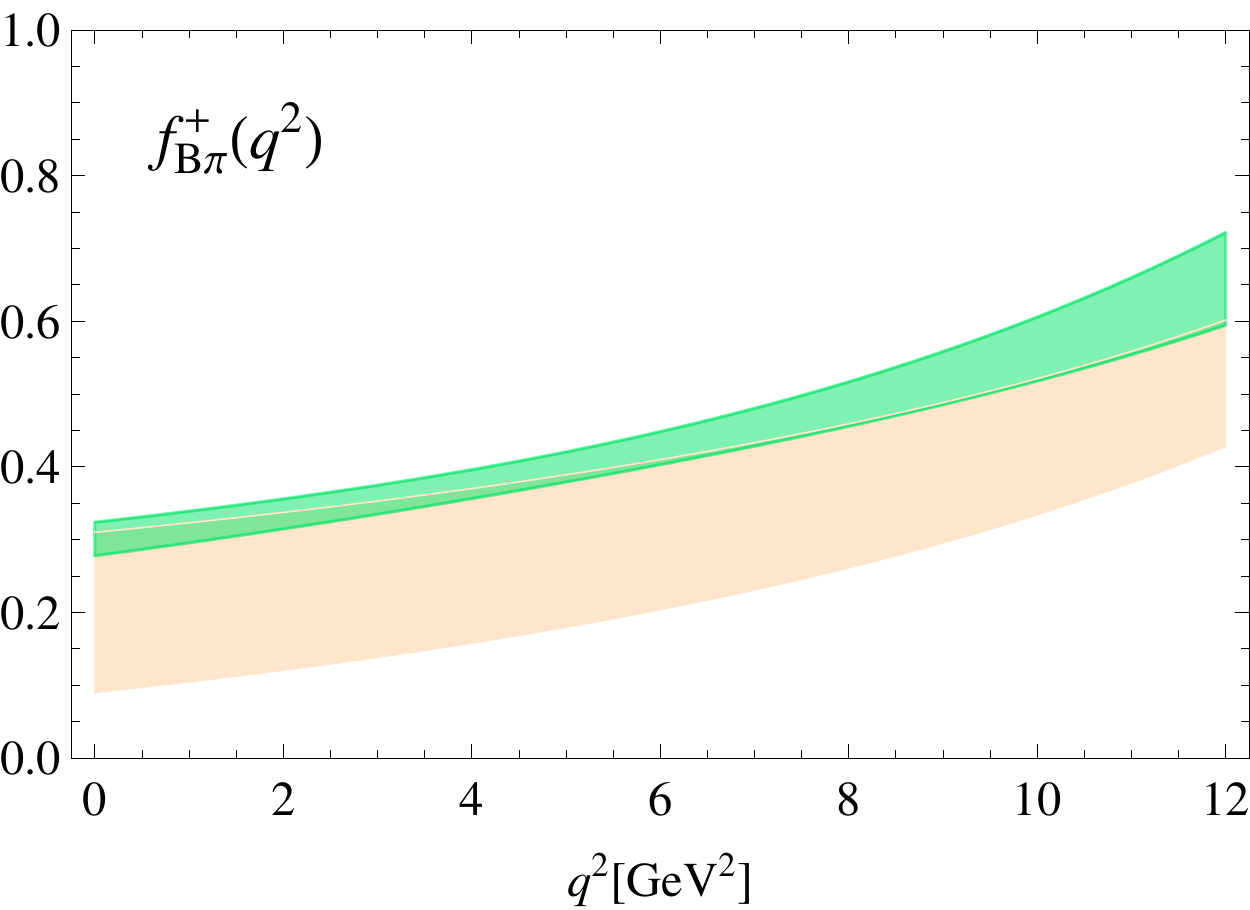} \quad
\includegraphics[scale=0.5]{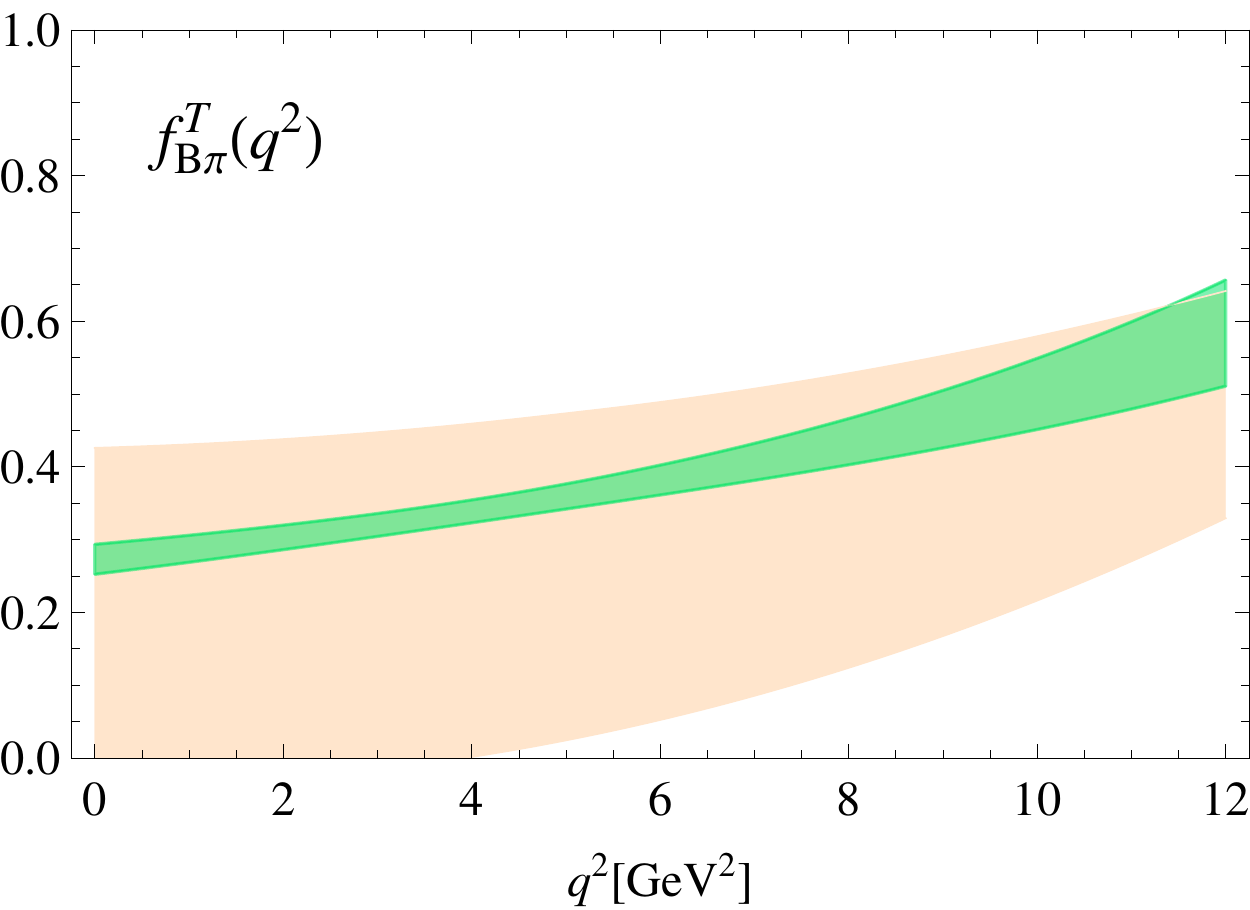}\\[2mm]
\includegraphics[scale=0.5]{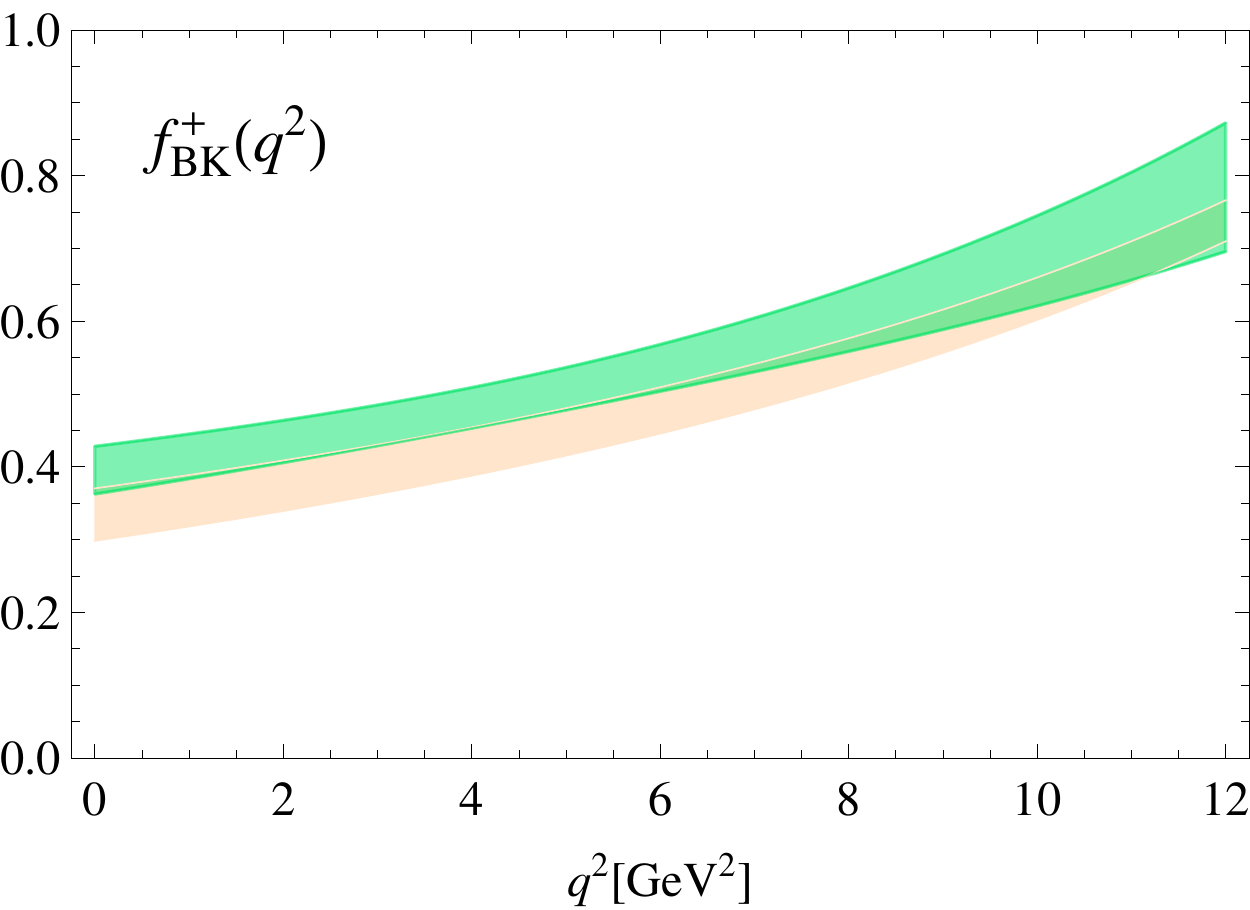} \quad
\includegraphics[scale=0.5]{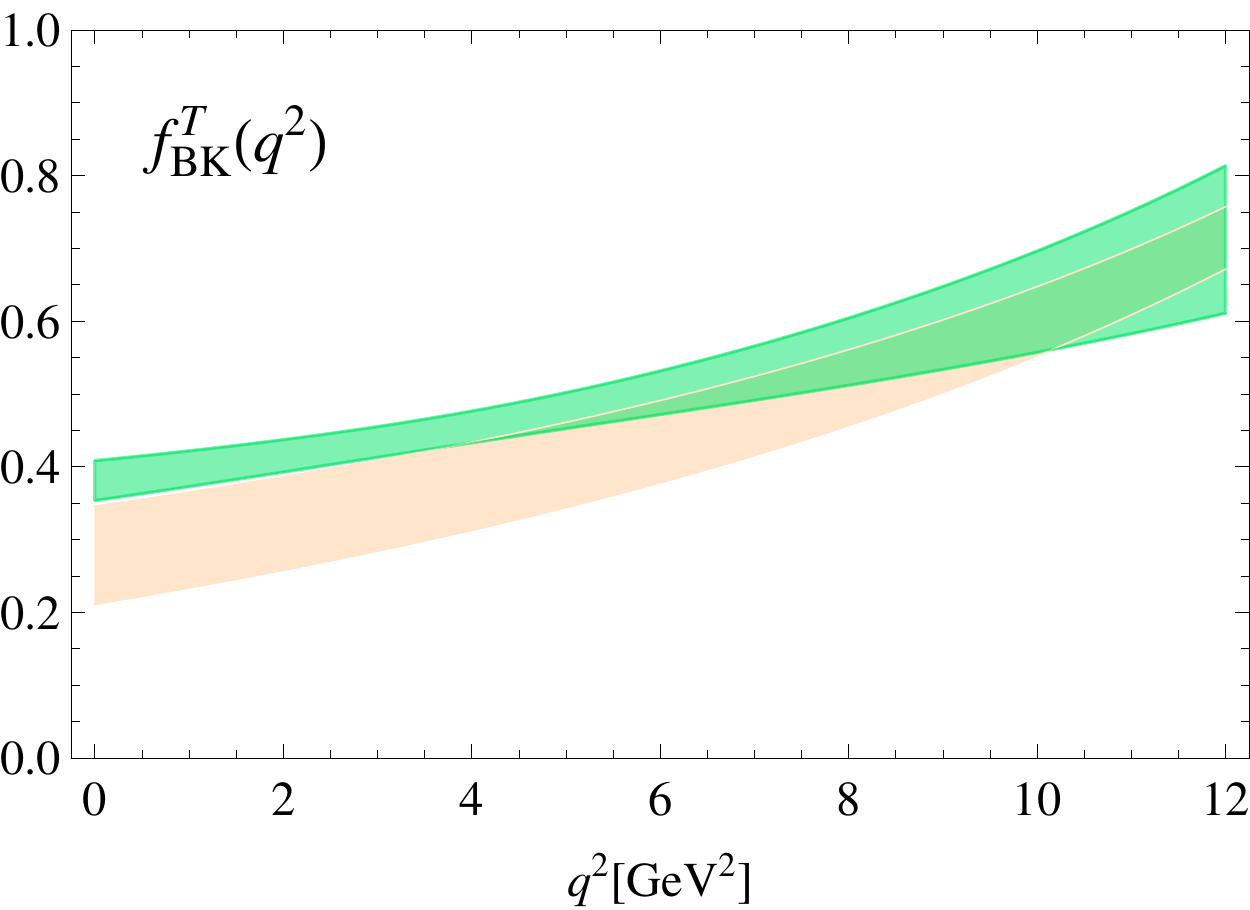}
\caption{ The vector (tensor) 
form factors of $B_s\to K$, $B\to K$ 
and $B \to \pi$ transitions 
calculated from  LCSRs including estimated parametrical 
uncertainties  are shown on the upper, middle
and lower left (right) panels, respectively, with the dark-shaded 
(green) bands. Extrapolations of the lattice QCD results 
for $B_s \to K$ \cite{Bouchard:2014ypa},  $B \to K$ \cite{Bailey:2015dka} 
and $B \to \pi$ \cite{Lattice:2015tia,Bailey:2015nbd} form factors 
are shown with the light-shaded 
(orange) bands.}
\label{fig:FFs}
\end{figure}

Using the input described above, we obtain the updated prediction 
for the $B_s \to K$ vector and tensor form factors
in the region $0\leq q^2\leq 12.0$ GeV$^2$ where the OPE
for LCSRs  in the adopted approximation is reliable (see Appendix A). 
In parallel, we also recalculate the $B \to K$ and  $B \to \pi$ form factors.
For convenience, we fit the LCSR predictions 
for the $B\to P$ form factors in this region to the two-parameter 
BCL-version of $z$-expansion
\cite{BCL} in the form adopted in  \cite{Khodjamirian:2009ys}:
\begin{eqnarray}
    f^{+,T}_{BP}(q^2) = \frac{f^{+,T}_{BP}(0)}{1 - q^2/m_{B^*_{(s)}}^2}
\Bigg\{1 + b_{1 (BP)}^{+,T} \Bigg[z(q^2) - z(0)
+ \frac{1}{2} \Big( z(q^2)^2 - z(0)^2\Big)\Bigg] \Bigg\}\,,
\label{eq:BCL}
\end{eqnarray}
where 
\begin{equation}
z(q^2) = \frac{{\sqrt{t_{+}-q^2}-\sqrt{t_{+}-t_{0}}}}
{{\sqrt{t_{+}-q^2}+\sqrt{t_{+}-t_{0}}}} \, , 
\end{equation}
\begin{equation}
t_{\pm} = (m_B  \pm m_{P})^2,
~~~ t_{0} = (m_B + m_P) \cdot (\sqrt{m_B} - \sqrt{m_P})^2\,,
\end{equation}
and the pole mass in Eq.~(\ref{eq:BCL}) for  $B_s\to K$, $B\to \pi$
($B\to K$) form factors is equal to $m_{B^*}$ ($m_{B_s^*}$). 
\begin{table}[h]
\begin{center}
\begin{tabular}{|c|c|c|c|}
\hline
Transition & $f^+_{BP}(0)$ & $b^+_{1\,(BP)}$ & Correlation \\
\hline
$B_s \to K $ & $0.336 \pm 0.023$ & $ -2.53 \pm 1.17 $ & $ 0.79 $ \\
\hline
$B \to K $ & $0.395 \pm 0.033$ & $ -1.42 \pm 1.52 $ & $ 0.72 $ \\
\hline
$B \to \pi$ & $0.301 \pm 0.023$ & $ -1.72 \pm 1.14 $ & $ 0.74 $ \\[1mm]
\hline
\hline
Transition & $f^T_{BP}(0)$ & $b^T_{1\,(BP)}$ & Correlation \\
\hline
$B_s \to K $ & $0.320 \pm 0.019$ & $ -1.08 \pm 1.53 $ & $ 0.74 $ \\
\hline
$B \to K $ & $0.381 \pm 0.027$ & $ -0.87 \pm 1.72 $ & $ 0.75 $ \\
\hline
$B \to \pi$ & $0.273 \pm 0.021$ & $ -1.54 \pm 1.42 $ & $ 0.78 $ \\
\hline
\end{tabular}
\end{center}
\caption{The fitted parameters of the $z$-expansion (\ref{eq:BCL}) 
for  the vector (upper panel) and tensor (lower panel) $B\to P$ form factors 
at $0<q^2<12.0$ GeV$^2$ calculated from LCSRs.}
\label{tab:BsK}
\end{table}
The fitted parameters of the vector and tensor form factors
and their correlations are presented in Table~\ref{tab:BsK}.
Note that, adopting a more complicated  $z$-expansion with more slope parameters,
only insignificantly changes the quality of the fit, 
and reveals strong correlations  between these parameters. In any case the
actual form of parametrization does not play a role 
as soon as we stay within the $q^2$-region 
where the form factors are directly calculated from LCSRs. 
Our results for the form factors are also plotted in Fig.~\ref{fig:FFs}, 
where the  error bands correspond to the uncertainties of the fitted 
parameters shown in Table~\ref{tab:BsK}.
For comparison, we also show in the same figures 
the extrapolations of the recent lattice QCD results 
obtained at large $q^2$ (low hadronic recoil) and continued to the small
$q^2$ region using the $z$-series parametrization.
For the vector $B_s\to K$ form factor this extrapolation was 
obtained by HPQCD Collaboration \cite{Bouchard:2014ypa}. 
The same form factor was also calculated by ALPHA Collaboration 
\cite{Bahr:2016ayy} at a single large-$q^2$ value.
For the vector and tensor  $B\to K$  form factors we compare 
our results with the 
extrapolations obtained from Fermilab Lattice and MILC Collaboration 
results \cite{Bailey:2015dka}, to which 
the HPQCD Collaboration results \cite{Bouchard:2013pna} are very close 
(not shown here).   
Finally, the low-$q^2$ extrapolations of the lattice $B\to \pi$
vector and tensor form factors are taken from  
\cite{Lattice:2015tia} and \cite{Bailey:2015nbd}, respectively.

The knowledge of the $B_s\to K$ vector form factor at large recoil enables us 
to calculate the quantity defined in Eq.~(\ref{eq:zeta}). The result
\begin{equation}
\Delta\zeta_{B_s K}\,[0, 12 \, {\rm GeV}^2] = 
7.03^{+0.73}_{-0.69} \, {\rm ps}^{-1} 
\end{equation}  
can be directly used for $|V_{ub}|$ determination, provided the
differential width of $B_s\to K \ell \nu_\ell$ integrated over the same bin is measured.  
For comparison we recalculate the same   
quantity for $B\to \pi \ell \nu_\ell$:
\begin{equation}
\Delta\zeta_{B \pi}\,[0, 12 \, {\rm GeV}^2] = 
5.30^{+ 0.67}_{-0.63} \, {\rm ps}^{-1}\,,
\label{eq:deltzeta}
\end{equation}  
which is, as it should be,  very close to the interval
predicted in \cite{Imsong:2014oqa}. The latter interval is somewhat 
narrower than (\ref{eq:deltzeta}), reflecting   
the statistical (Bayesian) treatment applied in \cite{Imsong:2014oqa} which generally
produces less conservative errors. 
In the future, when sufficiently  
accurate data on $B_s\to K \ell \nu_\ell$ become available,
a global statistical  treatment of all $B \to P$ form factors is desirable.

Comparing our results in Table~\ref{tab:BsK} with 
the earlier LCSR calculation \cite{Duplancic:2008tk}  of the $B\to K$ and $B_s\to K$ 
form factors, we emphasize that, albeit the numerical results look close to  
ours, there are differences in the subleading twist-3,4 terms.
We follow Ref.~\cite{Khodjamirian:2009ys} where 
these terms have already been discussed and corrected. Also, as
compared to \cite{Duplancic:2008tk}, we use slightly different $B_{(s)}$ 
decay constants and twist-3 normalization parameter $\mu_K$. 

Furthermore, the interval for our updated result for the $B\to K$ vector  
form factor in Table~\ref{tab:BsK} lies somewhat above the previous 
LCSR prediction \cite{Khodjamirian:2010vf}, 
$f^+_{BK}(0)=0.34^{+0.05}_{-0.02}$, mainly due to the smaller 
value of $f_B$  from the two-point sum rule used here 
and due to the slightly smaller value of the effective threshold 
in LCSR used in \cite{Khodjamirian:2010vf}.
On the other hand, in the LCSR for $f^T_{BK}(q^2)$ 
some minor corrections, implemented here in the subleading twist-4 terms,
largely compensate the shift caused by the  $B$-decay constant,
so that our result in Table~\ref{tab:BsK} is close to  $f^T_{BK}(0)=0.39^{+0.05}_{-0.03}$
obtained in \cite{Khodjamirian:2010vf}.

Turning finally to the LCSR result for the 
vector $B \to \pi$ form factor, 
which was updated several times in past, 
let us mention that although we use the same analytical expressions
as in Ref.~\cite{Duplancic:2008ix}, the input parameters such as $\mu_\pi$ 
(determined by the light quark masses) and Gegenbauer moments $a^\pi_2,a^\pi_4$ 
became  more accurate, leading 
to a narrower interval of our prediction, compared to 
the interval $f^+_{B\pi}(0)=0.26^{+0.04}_{-0.03}$ 
obtained in  Ref.~\cite{Duplancic:2008ix}. The central 
value of the latter is  somewhat below
the one we present in Table~\ref{tab:BsK}, since 
we use a smaller (larger) central input value 
of $f_B$ (of $\mu_\pi$). 
In Ref.~\cite{Duplancic:2008ix} 
one can also find a detailed comparison with the LCSR $B\to \pi$ form factor
obtained earlier in Ref.~\cite{Ball:2004ye}.

We turn to the numerical analysis of $B\to P\ell^+\ell^-$ observables,
where the $B\to P$ form factors obtained above are used.  
We recalculate  the nonlocal amplitudes, following \cite{Khodjamirian:2012rm,Hambrock:2015wka}.
In Appendix B a brief  outline of the calculational 
method is given. Here we need some additional input parameters. 
The most important are: the inverse moment $\lambda_B$ of the  $B$-meson DA 
(we assume $\lambda_{B_s} = \lambda_B$) and the Borel and threshold parameters in  
$\pi, K$ channel in the LCSRs for the soft-gluon emission contributions. 
They are displayed in Table~\ref{tab:input}. The same input parameters 
for the pion, kaon and $B$-meson DAs as the ones given in  Table~\ref{tab:input}
serve as an input in the hard-gluon contributions for which we use the 
QCD factorization expressions \cite{Beneke:2001at} at spacelike $q^2$.

The effective FCNC Hamiltonian (see Eq.~(\ref{eq:Heff}) in Appendix B) 
is chosen as in \cite{Hambrock:2015wka} (see Table V there), 
with all Wilson coefficients $C_i$ taken at leading order in $\alpha_s$. 
This accuracy is sufficient for $C_{1-6},C_8^{eff}$
entering the nonlocal hadronic amplitudes, having in mind the 
overall accuracy of our method for these amplitudes. 
At the same time, the numerically large 
Wilson coefficients $C_9,C_{10}$ and $C_{7}^{eff}$ of the FCNC operators 
multiplying the factorizable parts of the decay amplitudes,  
have a noticeable impact on  the observables.  
Therefore, we adopt here the values of these coefficients at the next-to-leading order
in $\alpha_s$ (see Table~\ref{tab:ci}).
\begin{table}[h]
\begin{center}
\begin{tabular}{|c|c|c|c|}
\hline
Coefficent & $\mu=2.5$ GeV & $\mu=3.0$ GeV & $\mu=4.5$ GeV\\
\hline
$C_{7}^{eff}$  &-0.332 & -0.321 & -0.304\\
& (-0.356) & (-0.343) & (-0.316) \\
\hline
$C_9$ & 4.070 & 4.076 & 4.115 \\
& (4.514) & (4.462) & (4.293) \\
\hline
$C_{10}$ & -4.122 & -4.122 & -4.122 \\
& (-4.493) & (-4.493) & (-4.493) \\
\hline
\end{tabular}
\end{center}
\caption{Wilson coefficients of the FCNC operators at next-to-leading 
(leading) order in $\alpha_s$ used in our numerical analysis at various scales.} 
\label{tab:ci}
\end{table}

For completeness and future use, in Appendix B the 
numerical results 
for the separate nonlocal amplitudes ${\cal H}^{(u)}_{BP}$
and ${\cal H }^{(c)}_{BP}$ defined as in Eq.~(\ref{eq:corr})
are presented in Figs.~\ref{fig:HBK},~\ref{fig:HBpi},~\ref{fig:HBsK}. 
Combining these results with the 
form factors, we compute the quantities defined in  
Eq.~(\ref{eq:Brbin}) for a single 
bin $[q_1^2,q_2^2]= [1.0~\mbox{GeV}^2, 6.0~\mbox{GeV}^2]$ which
optimally covers the part of the large-recoil region. 
The results are collected in Table~\ref{tab:BKllamplsq}, where the  
(uncorrelated) uncertainties are obtained by adding in quadrature 
the individual variations due to changes of input parameters.

\begin{table}
\begin{center}
\begin{tabular}{|c|c|c|c|c|}
\hline
Decay mode
& ${\cal F}_{BP} [1.0,6.0]$ & ${\cal D}_{BP} [1.0,6.0]$
& ${\cal C}_{BP} [1.0,6.0]$ & ${\cal S}_{BP} [1.0,6.0]$ \\
\hline
$B^-\to K^- \ell^+\ell^-$ &
$75.0^{+10.5}_{-9.7} $ & --- & --- & --- \\
\hline
$B^-\to \pi^- \ell^+\ell^-$ &
$47.7^{+6.4}_{-5.9}$ & 
$16.1^{+2.8}_{-10.1}$ & 
$14.3^{+7.8}_{-5.8}$ &
$-9.8^{+7.1}_{-7.2}$ \\
\hline
$\bar{B}_s\to K^0 \ell^+\ell^-$ &
$61.0^{+7.0}_{-6.8}$ & 
$7.8^{+3.4}_{-2.5} $ &
$-12.9^{+2.4}_{-2.2}$ &
$-3.4^{+1.1}_{-2.6}$ \\
\hline
\end{tabular}
\end{center}
\caption{
The parts of the $B\to P\ell^+\ell^-$ amplitudes squared,
as defined in Eqs.~(\ref{eq:binF})-(\ref{eq:binS}), in the units [GeV$^3$],
for the bin $[1.0~\mbox{GeV}^2,6.0~\mbox{GeV}^2]$.}
\label{tab:BKllamplsq}
\end{table}

Note that the binned quantities  ${\cal F}_{BP}$
are not much sensitive to the magnitude of the nonlocal
amplitudes  ${\cal H}^{(c)}_{BP} (q^2)$,  which enter
the numerically subleading contributions to the coefficients 
$c_{BP}(q^2)$.  Hence,  the differences 
between ${\cal F}_{BK}$, ${\cal F}_{B\pi}$ and ${\cal F}_{B_s K}$ in 
Table~\ref{tab:BKllamplsq}
roughly reflect the ratios of the corresponding form factors.
On the other hand, the remaining binned quantities 
${\cal D}_{BP}$, ${\cal C}_{BP}$ and ${\cal S}_{B P}$
are essentially determined by the nonlocal effects in $B\to P\ell^+\ell^-$.
In particular, the large differences between 
${\cal D}_{B\pi}$, ${\cal C}_{B\pi}$, ${\cal S}_{B\pi}$
and  ${\cal D}_{B_sK}$, ${\cal C}_{B_sK}$, ${\cal S}_{B_sK}$ 
emerge mainly due to 
the enhancement of the weak annihilation mechanism in the 
nonlocal amplitude ${\cal H}^{(u)}_{B\pi} (q^2)$ for 
$B^- \to \pi^- \ell^+ \ell^-$  \cite{Hambrock:2015wka}.  
The same mechanism does not play a role in the amplitude 
${\cal H}^{(u)}_{B_sK} (q^2)$ contributing to 
$\bar{B}_s \to K^0 \ell^+ \ell^-$, due to a 
different quark content of the initial  $B_s$ meson, and 
due to a suppressed  combination of Wilson coefficients.

As shown in the previous section, the binned quantities  
${\cal F}_{BP}$, ${\cal D}_{BP}$, ${\cal C}_{BP}$,
${\cal S}_{BP}$ can in principle be used for an independent  
determination of the Wolfenstein parameters $A$, $\eta$ and  $\rho$
from the combination of observables measured in 
$B \to P \ell^+ \ell^-$ decays. The important role
in this determination  is played by the direct $CP$-asymmetry 
in $B \to \pi \ell^+ \ell^-$ which is not available yet in the 
large-recoil region bins. Hence, here we limit ourselves by an inverse 
procedure. Taking  the values  of all Wolfenstein  parameters 
\begin{eqnarray}
\lambda = 0.22506 \pm 0.00050,~~  
A = 0.811 \pm 0.026,~~
\nonumber\\
\bar \rho = \rho \left(1-\frac{\lambda^2}{2}\right)=0.124^{+0.019}_{-0.018} , 
~~\bar \eta = \eta \left(1-\frac{\lambda^2}{2}\right)= 0.356 \pm 0.011 \,, 
\label{eq:ckmfit}
\end{eqnarray}
from the global fit 
of CKM matrix \cite{pdg} and
using the calculated hadronic input from Table~\ref{tab:BKllamplsq},
we predict the values of the binned branching fractions  presented in 
Table~\ref{tab:BKllBR}  and the binned direct $CP$-asymmetries:
\begin{equation}
{\cal A}_{B\pi}[1.0,6.0]= -0.15^{+0.11}_{-0.11}\,,~~ ~~
{\cal A}_{B_sK}[1.0,6.0]=-0.04^{+0.01}_{-0.03}\,.
\label{cpres}
\end{equation}

\begin{table}[t]
\begin{center}
\begin{tabular}{|c|c|c|c|}
\hline
Decay mode & $B^-\to K^- \ell^+\ell^-$ & $B^-\to \pi^- \ell^+\ell^-$  &  
$\bar{B}_s\to K^0 \ell^+\ell^-$ \\ 
\hline
Measurement 
&${\cal B}_{BK}[1.0\,,6.0]$ &${\cal B}_{B\pi}[1.0\,,6.0]$&
${\cal B}_{B_sK}[1.0,6.0]$\\[-2mm]
or calculation &&&\\
\hline
Belle \cite{Wei:2009zv}&$2.72\,^{+0.46}_{-0.42}\pm\, 0.16 $&---&---\\
\hline
CDF \cite{Aaltonen:2011qs} &2.58\,$\pm$\,0.36$\,\pm$\,0.16&---&---\\
\hline
BaBar \cite{Lees:2012tva}&$\!2.72\,^{+0.54}_{-0.48}\pm 0.06 $&---&---\\
\hline
LHCb \cite{Aaij:2014pli}, \cite{LHCb:Bpimumu15}  &$2.42 \pm 0.7\pm 0.12$\,
                                       &$0.091^{+0.021}_{-0.020} \pm
                                         0.003$\,&---\\
\hline
\hline
HPQCD \cite{Bouchard:2013mia}&$3.62 \pm 1.22$&---&---\\
\hline
Fermilab/MILC \cite{Bailey:2015nbd}, \cite{Du:2015tda}  
& $3.49 \pm 0.62$  & $ 0.096 \pm 0.013$ & --- \\
\hline
\hline
This work &
$ 4.38^{+0.62}_{-0.57} \pm 0.28$ &
$ 0.131^{+0.023}_{-0.022} \pm 0.010 $ &
$ 0.154^{+0.018}_{-0.017} \pm 0.011 $ \\
\hline
\end{tabular}
\end{center}
\caption{Binned branching fractions in the units of
$10^{-8}~\mbox{GeV}^{-2}$ 
defined in Eq.~(\ref{eq:Brbinav}) for the bin  
$[q^2_1,q_2^2]=[1.0~{\rm GeV}^2- 6.0~{\rm GeV}^2] \, $. 
The first (second) error in our predictions is due to the uncertainty
of the input (only of the CKM parameters).}
\label{tab:BKllBR}
\end{table}
The numerical results for the $B \to K \ell^+ \ell^-$ 
and $B\to \pi\ell^+\ell^-$ decays presented here
update the previous ones obtained, respectively, 
in \cite{Khodjamirian:2012rm} 
\footnote{Note that the branching
fractions given in the literature are adjusted to our definition,
which implies division by the width $(q_2^2-q_1^2)$ of the bin}
and \cite{Hambrock:2015wka}.
 
\section{Discussion}

In this paper we updated the LCSR predictions for the 
$B_s\to K$ form factors in the large recoil region of the kaon. 
We predicted the ratio of the integrated $B_s\to K \ell \nu_\ell$ decay width and $|V_{ub}|^2$.  Our result can be used to determine this CKM matrix element 
from the future data on $B_s\to K \ell \nu_\ell$ in the kinematically
dominant large recoil region.

We also calculated the hadronic input for the branching fractions and 
direct $CP$-asymmetries of $B\to P \ell^+\ell^-$  FCNC decays
in the large recoil bin  $ 1.0\leq q^2 \leq 6.0$ GeV$^2$. 
Our results include the  $B\to P$  form factors and nonlocal 
hadronic matrix elements, all obtained in the same framework
and with a uniform input. The LCSRs used in this calculation 
take into account the soft-overlap nonfactorizable contributions to the form factors
and nonlocal amplitudes.  
Extending the application of LCSRs to other 
nonlocal contributions represents an important task for the future.
For example, as discussed in more detail in \cite{Hambrock:2015wka}, 
the weak annihilation contribution which is important in the
$B\to \pi \ell^+\ell^-$ decay can be obtained from LCSRs with $B$-meson DAs,
alternative to QCD factorization and potentially including subleading effects.

Furthermore, we suggested a systematic way  
to extract the CKM matrix elements, cast in a form of   
the Wolfenstein parameters, from 
the combination of observables in $B\to P \ell^+\ell^-$  decays,
independent of the other methods involving the nonleptonic 
$B$-decays and/or $B-\bar{B}$ mixing.

Note that an independent extraction of CKM parameters 
is also possible from other modes of FCNC exclusive $B$-decays, 
such as $B_{(s)}\to V\gamma$ or $B_{(s)}\to V\ell^+\ell^-$, where $V=K^*, \rho$. 
The corresponding combinations of observables demand, apart from 
$B\to V$ form factors, a dedicated calculation of
all  relevant nonlocal hadronic matrix
elements. For this not yet accomplished task, a variety of methods 
combining QCD factorization with various versions of LCSRs 
may prove to be useful. In case of radiative decays the sum rules
with photon and vector-meson DAs and heavy-meson interpolating currents 
can be also of use (for previous works in this direction see 
\cite{Khodjamirian:1995uc,Ali:1995uy,Lyon:2013gba}).

In Table~\ref{tab:BKllBR} we compare our results 
for the binned branching fractions $^\arabic{footnote}$
with the experimental measurements and 
lattice QCD predictions \cite{Bouchard:2013mia,Bailey:2015nbd,Du:2015tda}. 
In the lattice QCD studies of $B\to P\ell^+\ell^-$ decays, as explained in detail 
in \cite{Du:2015tda}, the nonlocal contributions cannot be  calculated 
in a fully model-independent way. Instead, the (continuum) QCD-factorization 
\cite{Beneke:2001at} in the timelike region of $q^2$ is employed. 
Let us also mention in this context the earlier estimates of 
$B\to K \ell\ell$ \cite{Bobeth:2012vn,Altmannshofer:2012az}
and $B \to \pi \ell \ell$ \cite{Ali:2013zfa}
where the QCD-factorization approach was used combined with various inputs 
and extrapolations for the form factors.  

As seen from Table~\ref{tab:BKllBR}, the theory predictions for the $B\to K \ell^+
\ell^-$ branching fraction reveal some tension with the experimentally measured 
values, making this observable an important ingredient of the global fits of 
rare $B$ decays (see e.g., \cite{Descotes-Genon:2015uva}). 
Adding the characteristics of $B\to \pi\ell^+\ell^-$ and 
$B_s\to K\ell^+\ell^-$ decays to the set of 
fitted observables will further extend the possibilities to test the 
Standard Model in the quark-flavour sector. The fact that these very rare  
$B$-decay modes are within the reach of LHCb experiment, makes this task realistic.

\section*{Acknowledgments}
We thank Danny van Dyk and Yu-Ming Wang for useful discussions. 
This work is supported by the DFG Research Unit FOR 1873 "Quark Flavour Physics and Effective Theories", contract No KH 205/2-2.
AK is grateful for support to the Munich Institute for Astro- and Particle Physics (MIAPP) of the DFG cluster of excellence "Origin and Structure of the Universe" where the part of this work was done.
AR acknowledges the Nikolai-Uraltsev Fellowship of Siegen University 
and the partial support of the Russian Foundation for 
Basic Research (project No. 15-02-06033-a).

\appendix

\section{LCSR calculation of the {\boldmath $B\to P$} form factors}
The LCSRs for $B\to P$ ($P=\pi,K$) form factors at large recoil of $P$
(parametrically, at $q^2\ll m_b^2$) are derived from the 
correlation function of the weak flavour-changing current  
and $B$-interpolating quark current, sandwiched between the vacuum and 
on-shell $P$-state: 
\begin{eqnarray}
F^\mu_{BP} (p, q) & = & i \int d^4 x e^{i q x} 
\langle P (p) | T\{\bar q_1 (x) \Gamma^\mu b(x), (m_b+m_{q_2}) \bar b(0) i \gamma_5 q_2(0)\} 
|0 \rangle \nonumber \\
& = &  
\left\{ 
\begin{array}{ll}
F_{BP} (q^2, (p+q)^2) p^\mu + \tilde{F}_{BP} (q^2, (p+q)^2) q^\mu, 
& \quad \Gamma^\mu = \gamma^\mu\,, \\
F^T_{BP} (q^2, (p+q)^2) \left[q^2 p^\mu  - (q \cdot p) q^\mu 
\right], & \quad \Gamma^\mu = - i \sigma^{\mu \nu} q_\nu\,,
\end{array}
\right.
\label{eq:corr-func-def}
\end{eqnarray}
where the quark-flavour combination $q_1=u$, $q_2=s$  
corresponds to the $\bar{B}_s\to K^+$ weak transition;  
$q_1=s$, $q_2=u$  and  $q_1=d$ and $q_2=u$ ($q_2=s$)  correspond,
respectively to the $B^-\to K^-$  and $B^-\to \pi^-$  ($\bar{B}_s\to K^0$) FCNC transitions. 

The invariant amplitudes $F_{BP} (q^2, (p+q)^2)$ and $F^T_{BP} (q^2, (p+q)^2)$
in (\ref{eq:corr-func-def})
are used to derive the LCSRs for
the vector $f_{BP}^+ (q^2)$ and  tensor $f_{BP}^T (q^2)$ form factors, respectively.
At $q^2 \ll m_b^2$ and $(p+q)^2 \ll m_b^2$  
the OPE near the light-cone $x^2\simeq 0$ is applied for the correlation
function (\ref{eq:corr-func-def})
and the result is cast in a form of convolution, e.g.,:
\begin{equation}
F^{\rm (OPE)}_{BP} (q^2, (p+q)^2) = \!\!\! \!\!\sum_{t = 2,3,4,\ldots} \!\!\!\!\!\int {\cal D} u \!\!\!\sum_{k = 0,1,...} 
\!\!\!\!\!\left(\frac{\alpha_{s}(\mu)}{\pi}\right)^k\!\! T_k^{(t)} (q^2, (p+q)^2, \{u_i\}) \varphi_P^{(t)} (\{u_i\},\mu), 
\label{eq:OPE}
\end{equation}
where $T_k^{(t)}$ are the perturbatively calculable hard-scattering 
amplitudes and $\varphi^{(t)}_P ({u_i})$ are the 
$P$-meson light-cone distribution amplitudes (DAs) of the twist $t\geq 2$. 
The variables $\{u_i\} = \{u_1, u_2, ...\}$ are the fractions of the $P$-meson 
momentum carried by the constituents of DAs 
and ${\cal D} u = \delta (1 - \sum_i u_i) \prod_i d u_i$.
In Eq.~(\ref{eq:OPE}) the same renormalization 
scale  $\mu$ is used for DAs and for the
QCD running parameters in the adopted $\overline{MS}$ scheme.

The terms in the Eq.~(\ref{eq:OPE}) that correspond to higher-twist light meson 
DAs are suppressed by inverse powers of the $b$-quark virtuality 
$\sim((p+q)^2 - m_b^2) \sim \bar \Lambda m_b$, where $\bar \Lambda \gg \Lambda_{\rm QCD}$
does not scale with $m_b$. The adopted approximation for the correlation function
includes LO contributions of the twist 2,3,4 quark-antiquark
and quark-antiquark-gluon DAs. For the kaon DAs the $O(m^2_K)\sim O(m_s)$ accuracy 
is adopted.
The factorizable parts of twist-5,6 
contributions to LCSRs for $B \to P$ form factors 
were calculated by one of us \cite{Rusov_tw56} 
and their numerical impact on the total invariant amplitude was 
found negligible, $ < 0.1 \%$ of the total. This  
strengthens the argument for using a truncated twist expansion
to the accuracy $t=4$.

The NLO $O (\alpha_s)$ corrections to the twist-2 
and (two-particle) twist-3 hard-scattering amplitudes $T^{(2,3)}_1$
are taken into account. In the latter we neglect the $s$-quark mass,
hence, the double suppressed $O(\alpha_s m_s/\bar \Lambda)$ effects.
We use the  expressions for OPE derived in \cite{Duplancic:2008ix} 
extending them to the $B\to K$ and $B_s\to K$ cases (see also \cite{Khodjamirian:2009ys}). 
We do not include the $O(\beta_0)$ estimate of the twist-2 
$O (\alpha_s^2)$ contribution to the twist-2 hard-scattering 
amplitude calculated in \cite{Bharucha:2012wy}, since the resulting
effect in LSCR is very small and does not yet represent a 
complete NNLO computation of $T_1^{(2)}$.

The analytic result for $F^{\rm (OPE)}_{BP}(q^2,(p+q)^2)$ and 
$F^{T{\rm (OPE)}}_{BP}(q^2,(p+q)^2)$  is matched to the hadronic dispersion relation 
for the correlation function (\ref{eq:corr-func-def}) in the variable $(p+q)^2$.
To apply quark-hadron duality one  needs to transform the calculated
invariant amplitudes to the form of dispersion integral, 
\begin{eqnarray}
F^{(T){\rm (OPE)}}_{BP}(q^2,(p+q)^2)=
\frac1{\pi}\int\limits_{m_b^2}^{\infty} ds \frac{\mbox{Im}F^{(T){\rm (OPE)}}_{BP}(q^2,s)}{s-(p+q)^2}\,.
\label{eq:dispOPE}
\end{eqnarray}
We equate the contribution of the excited and continuum 
$B$-states in the hadronic dispersion relation to
the part of the above integral at $s>s_0^B$, 
where $s_0^B$ is the effective, process-dependent threshold.
The integral at $s\leq s_0^B$ is then 
equated to the contribution of the ground-state of $B$-meson.
The subsequent Borel transformation with respect 
to the variable  $(p+q)^2$  exponentiates
denominators, so that, e.g., $1/[s-(p+q)^2] \to e^{-s/M^2}$. 
Here $M^2$ is the Borel parameter chosen so that 
$M^2\sim \overline{\Lambda}m_b\sim \mu^2$ 
guarantees a  power suppression of higher-twist 
contributions. One finally obtains the LCSRs 
for the $B \to P$ form factors:
\begin{eqnarray}
f_{BP}^+ (q^2) & = &  \frac{e^{m_B^2/M^2}}{2 m_B^2 f_B} 
\frac1{\pi}\int\limits_{m_b^2}^{s_0^B} ds~ 
\mbox{Im}F^{\rm (OPE)}_{BP}(q^2,s) e^{-s/M^2}\,,
\nonumber \\
f_{BP}^T (q^2) & = & \frac{(m_B + m_P) e^{m_B^2/M^2}}{2 m_B^2 f_B}
\frac1{\pi}\int\limits_{m_b^2}^{s_0^B} ds~ 
\mbox{Im}F^{T {\rm (OPE)}}_{BP}(q^2,s) e^{-s/M^2}\,.
\label{eq:lcsrs}
\end{eqnarray}

\section{Nonlocal contributions to {\boldmath $B \to P \ell^+ \ell^-$}}

\begin{figure}[t]\center
\includegraphics[scale=0.5]{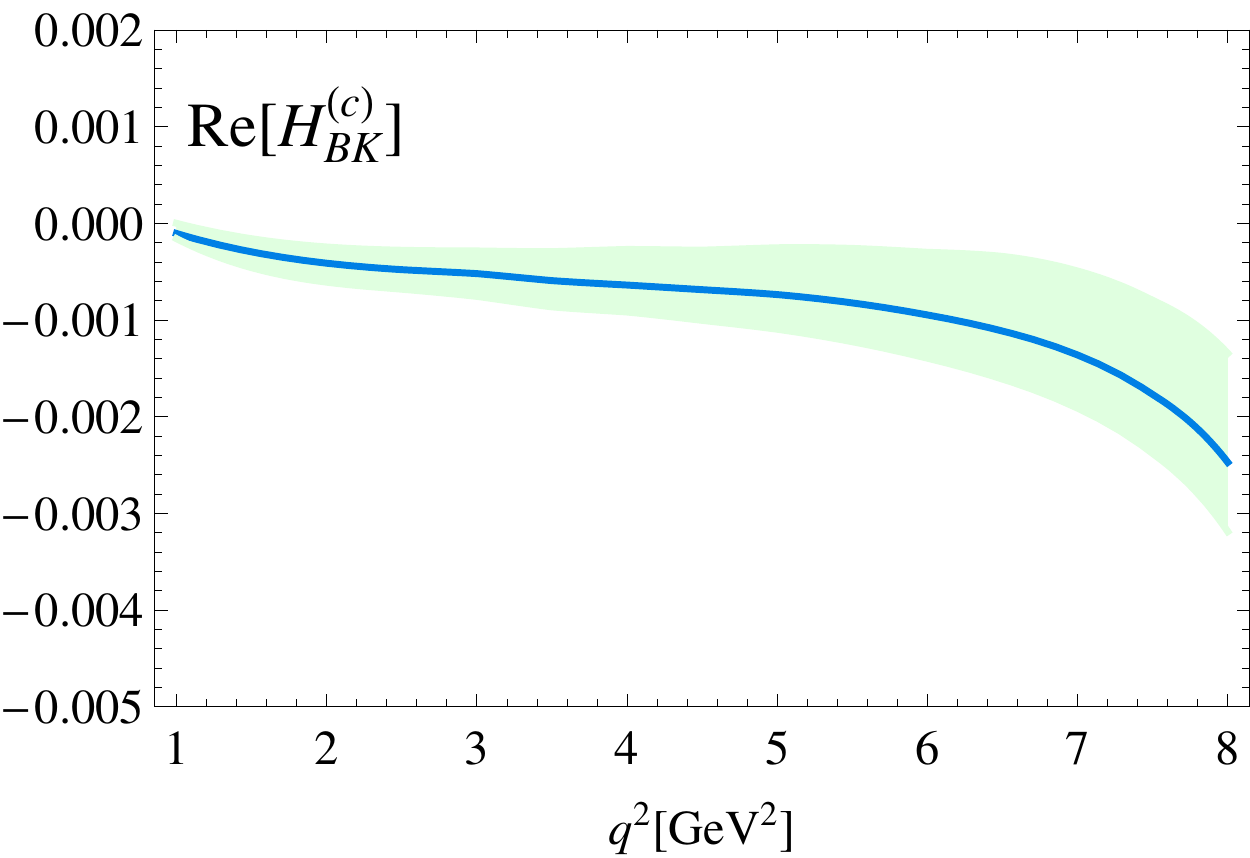}
\includegraphics[scale=0.5]{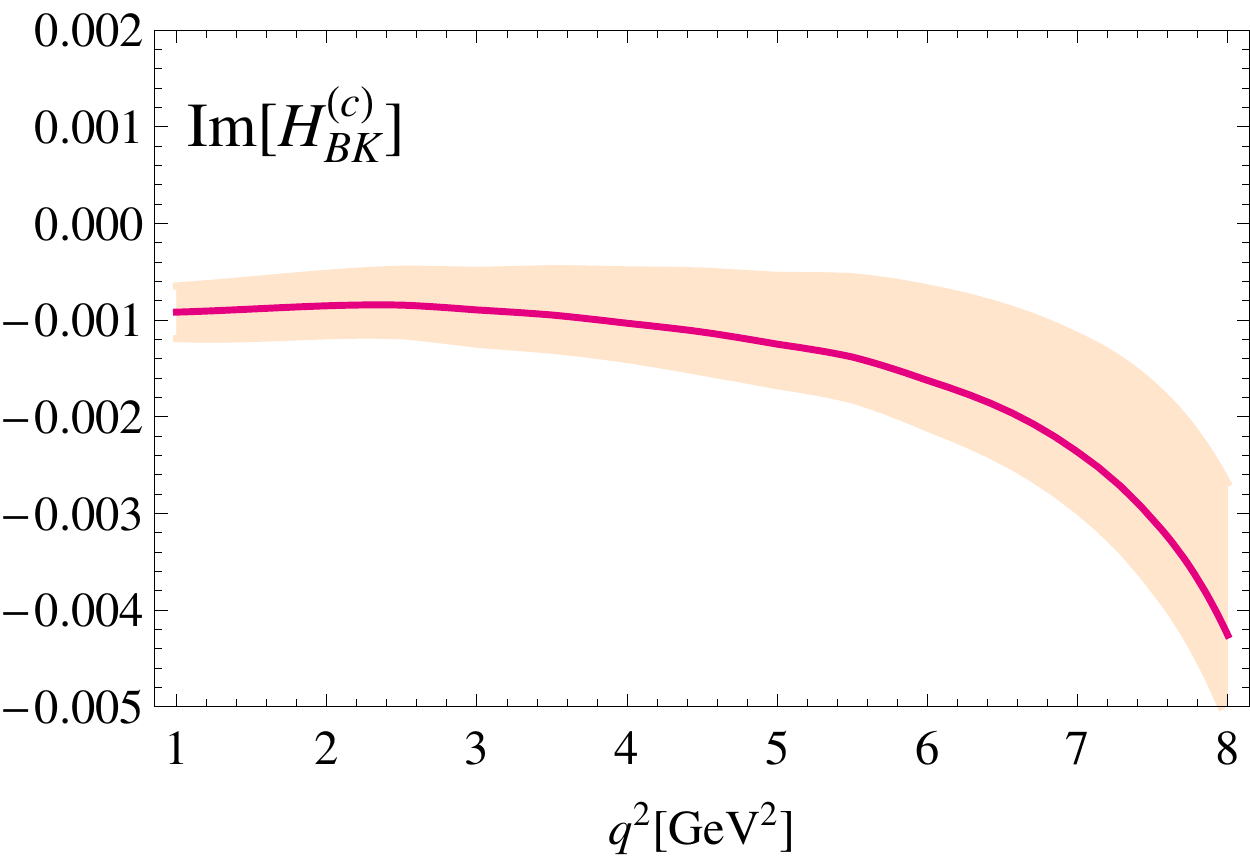}
\caption{Hadronic nonlocal amplitude ${\cal H}_{BK}^{(c)} (q^2)$
in $B^- \to K^- \ell^+ \ell^-$ in the large recoil region. 
On the left (right) panel the real (imaginary) part is plotted for the central input (solid) 
and including uncertainties (dashed band).}
\label{fig:HBK}
\end{figure}

\begin{figure}[ht]\center
\includegraphics[scale=0.5]{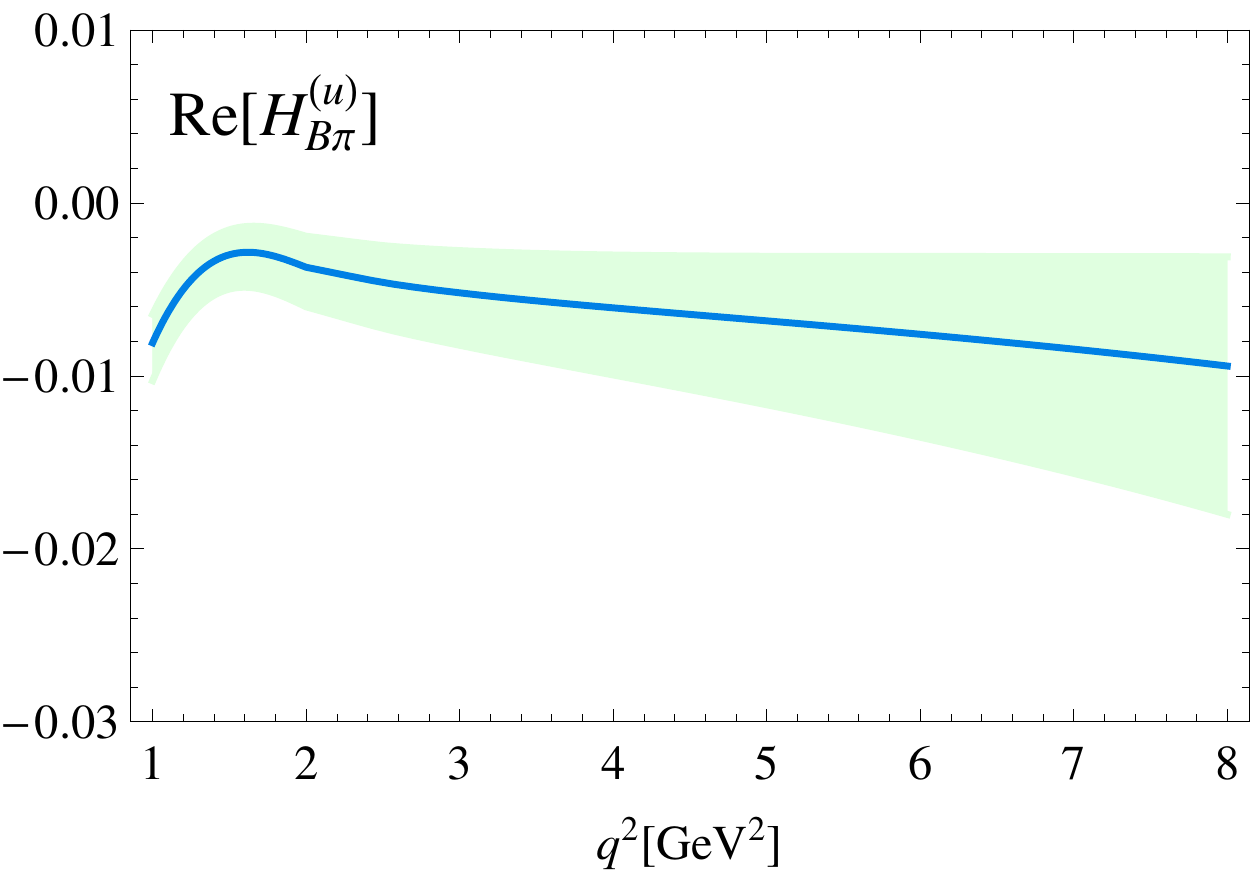}
\includegraphics[scale=0.5]{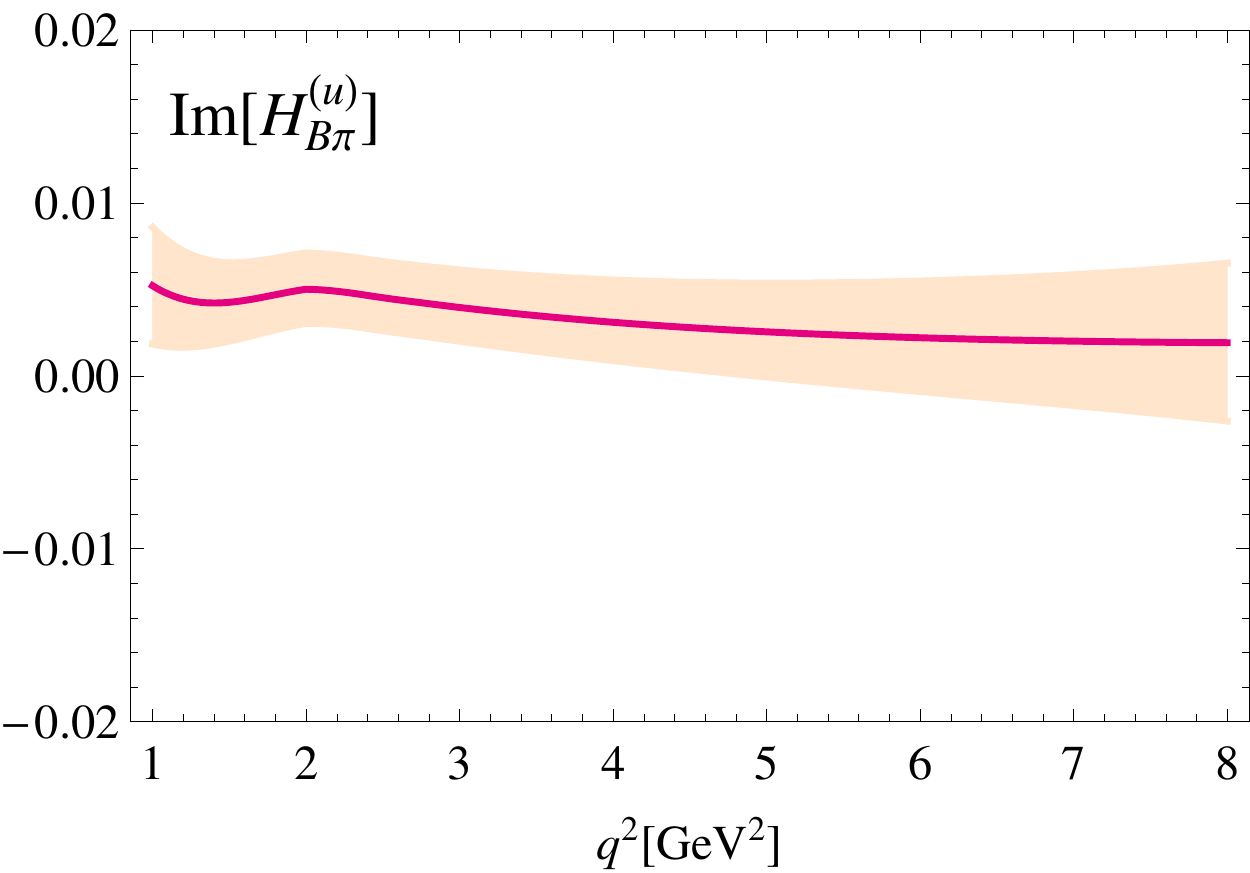}
\includegraphics[scale=0.5]{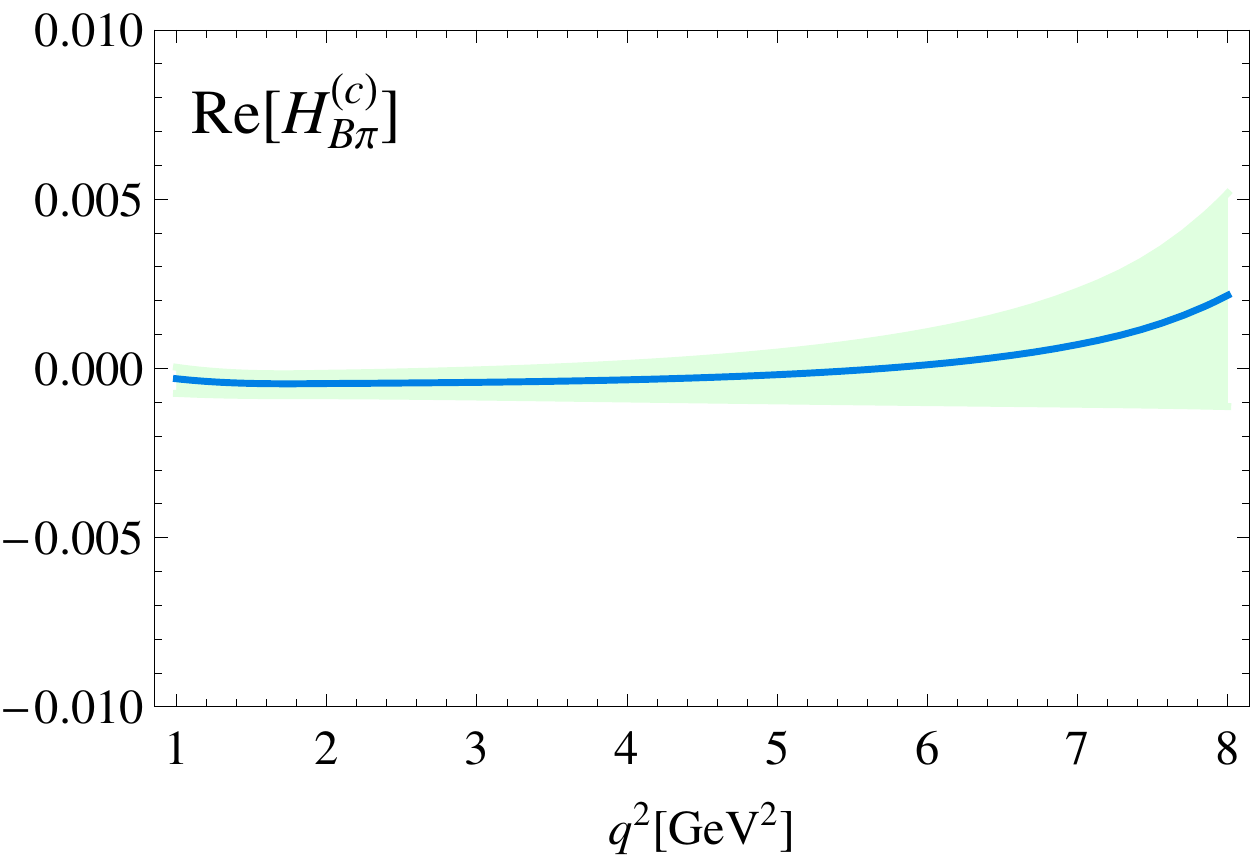}
\includegraphics[scale=0.5]{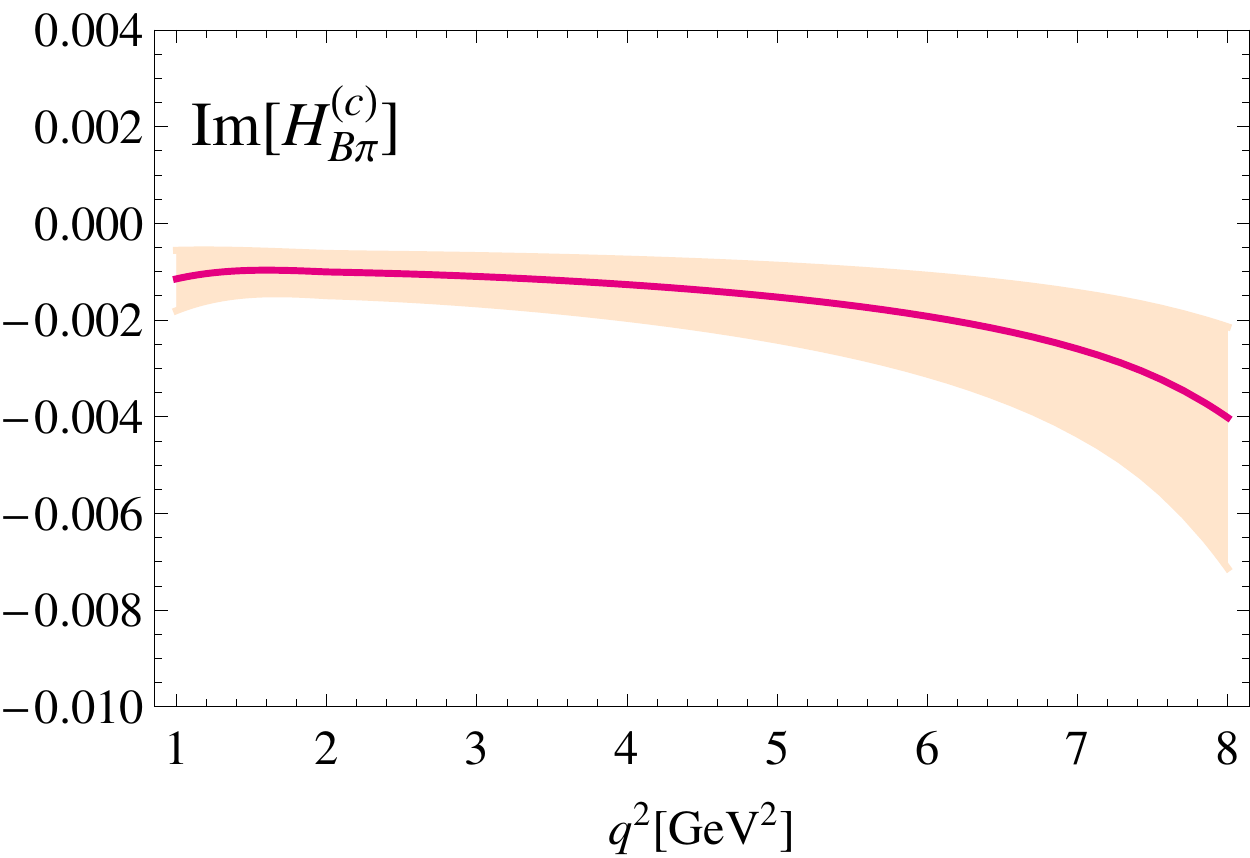}
\caption{The same as in Fig.~\ref{fig:HBK} 
for the amplitudes ${\cal H}_{B\pi}^{(u)} (q^2)$
and ${\cal H}_{B\pi}^{(c)} (q^2)$ in 
$B^- \to \pi^- \ell^+ \ell^-$.}
\label{fig:HBpi}
\end{figure}

\begin{figure}[ht]\center
\includegraphics[scale=0.5]{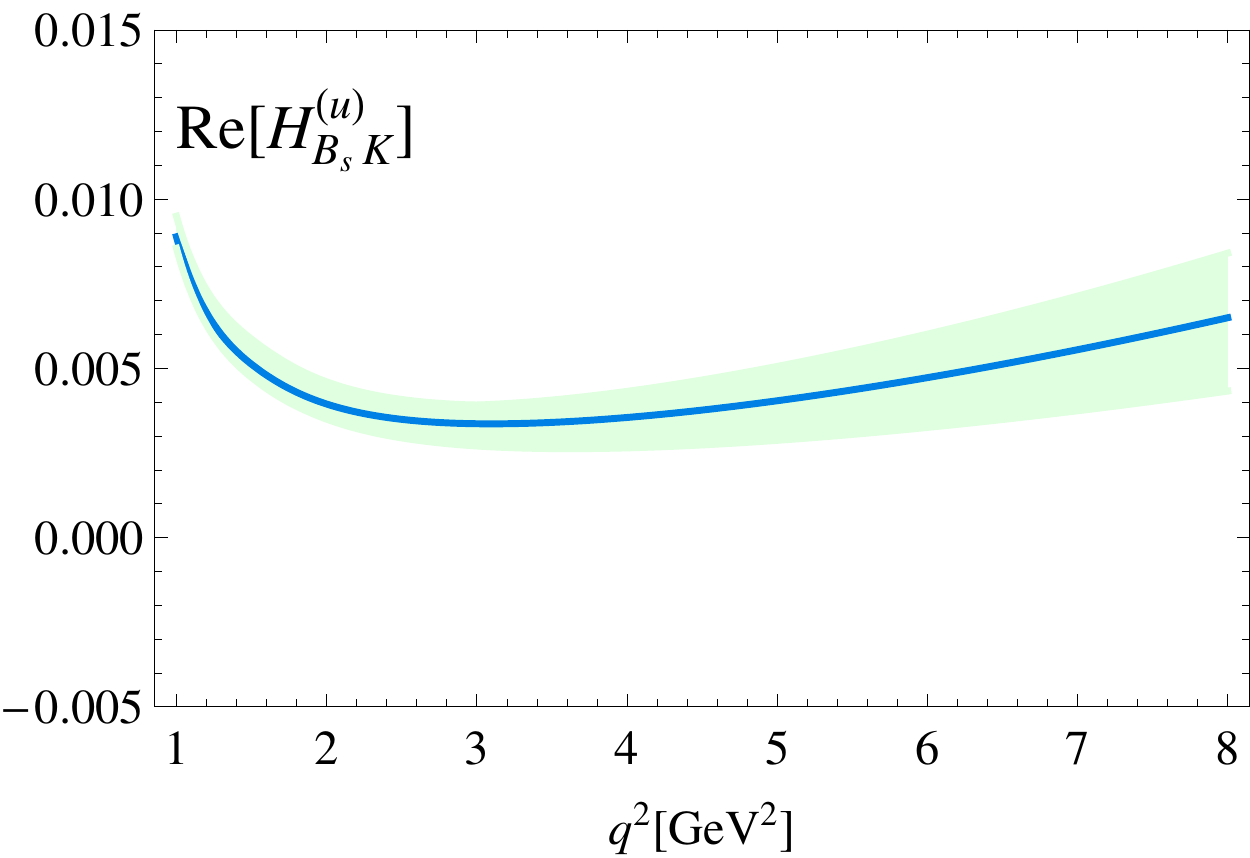}
\includegraphics[scale=0.5]{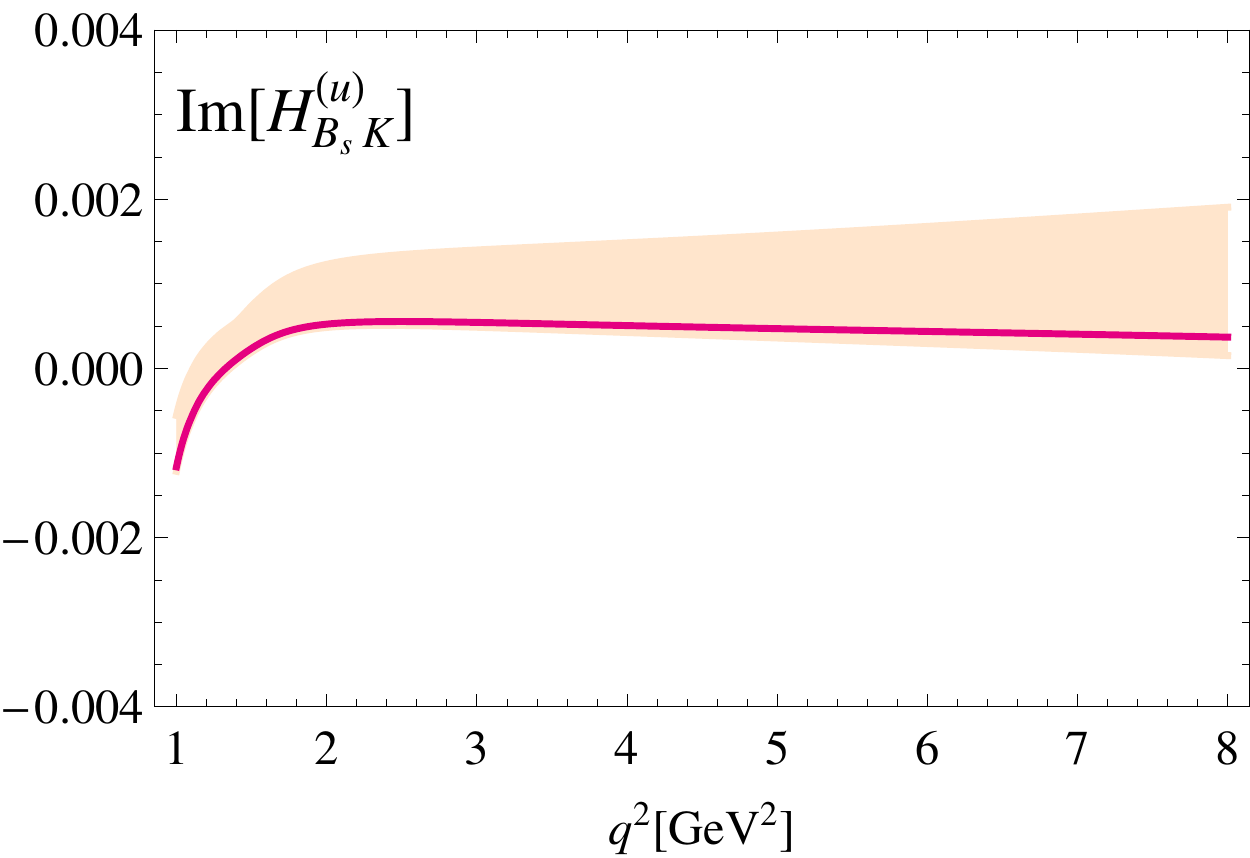}
\includegraphics[scale=0.5]{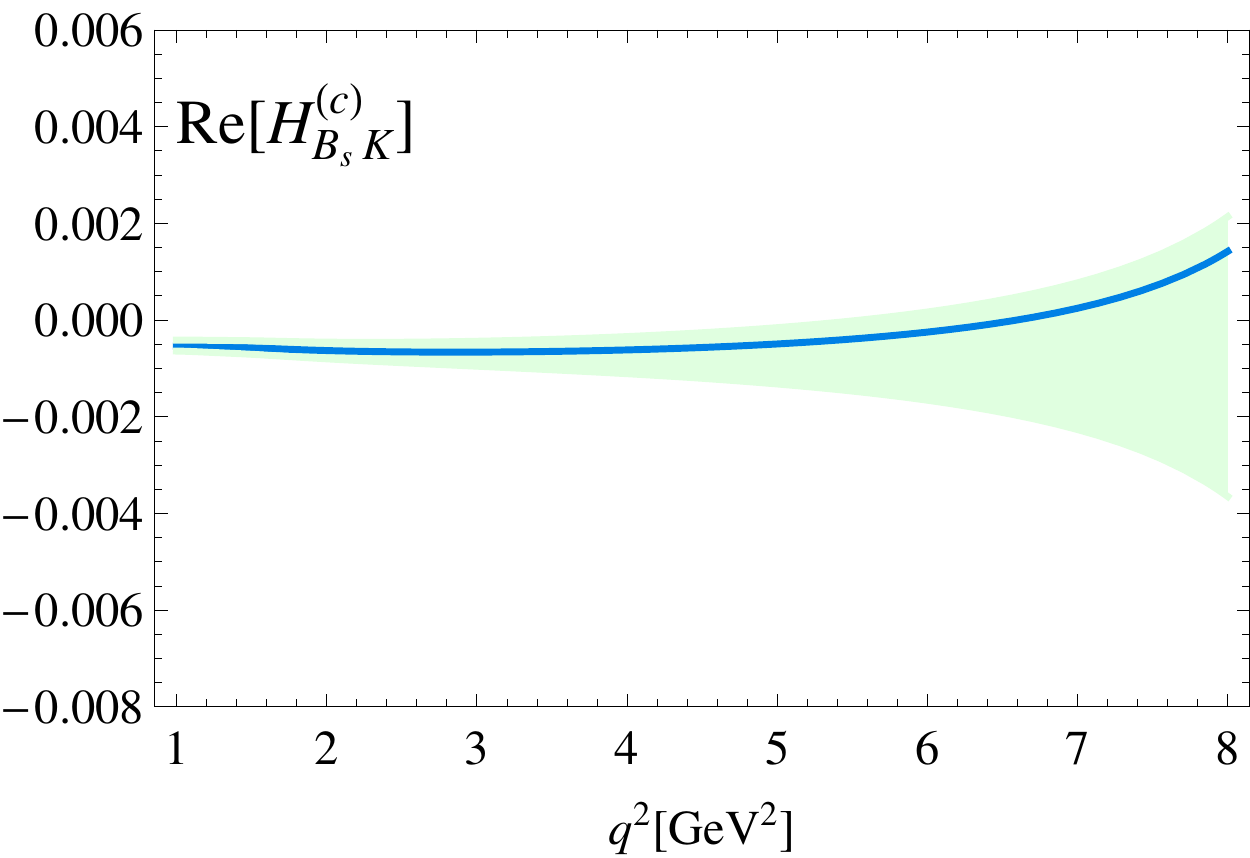}
\includegraphics[scale=0.5]{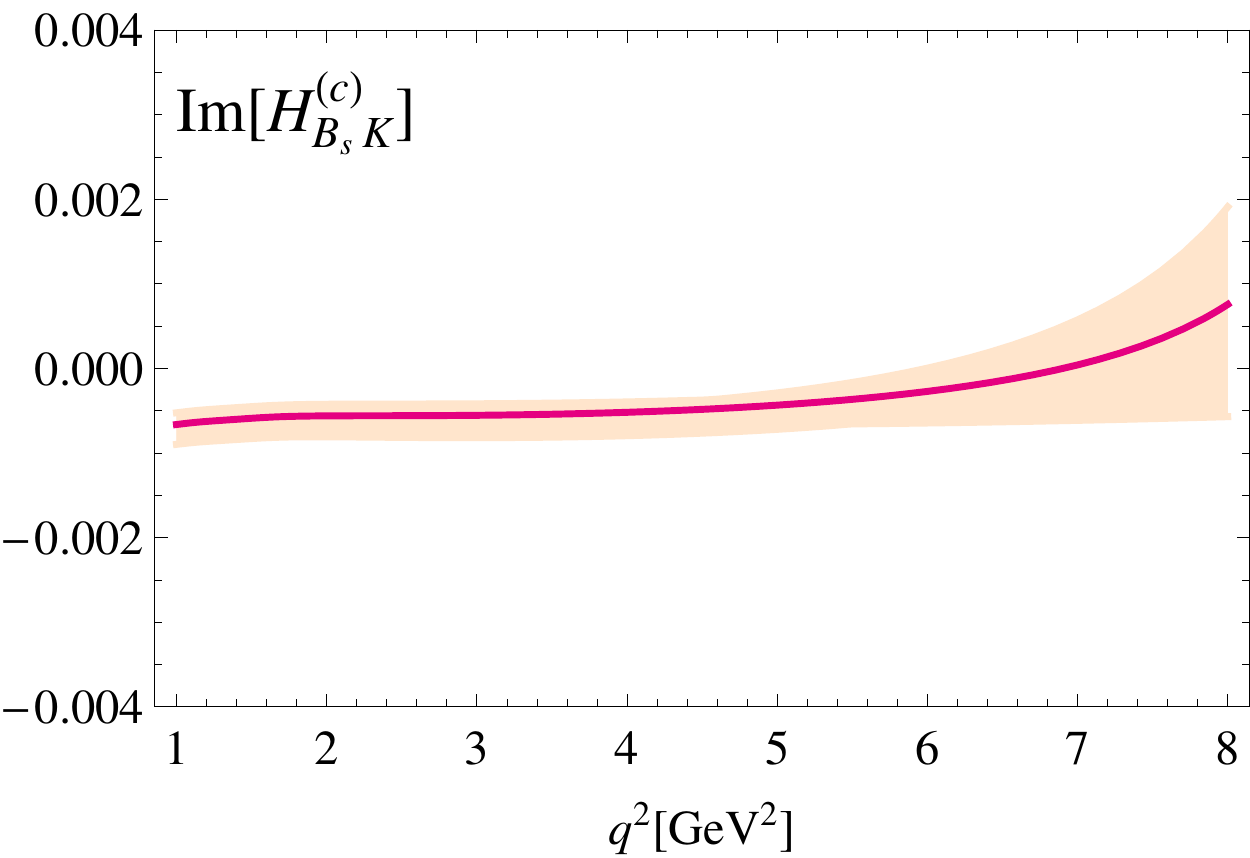}
\caption{The same as in Fig.~\ref{fig:HBK} 
 for the amplitudes ${\cal H}_{B_s K}^{(u)} (q^2)$
and ${\cal H}_{B_sK}^{(c)} (q^2)$ in 
$\bar{B}_s \to K^0 \ell^+ \ell^-$. }
\label{fig:HBsK}
\end{figure}

The effective weak Hamiltonian of the $b \to q \ell^+ \ell^-$  transitions 
($q = d,s$) generating  the $B \to P \ell^+ \ell^-$  decays  
has the following form in the Standard Model (see e.g., the review~\cite{Buchalla:1995v}):
\begin{equation}
\label{eq:Heff}
H^{b \to q}_{\rm eff} = 
\frac{4 G_F}{\sqrt 2} \left(
\lambda_u^{(q)} \sum\limits_{i=1}^{2} C_i \, {\cal O}_i^{u} 
+\lambda_c^{(q)} \sum\limits_{i=1}^{2} C_i \, {\cal O}_i^{c}
-\lambda_t^{(q)} \sum\limits_{i=3}^{10} C_i \, {\cal O}_i 
\right) + h.c.\,,
\end{equation}
where $\lambda_p^{(q)} = V_{pb} V_{pq}^*$, ($p = u,c,t$)
are the products of CKM matrix elements.
For the $B\to K \ell^+\ell^-$ transitions, the 
part of the decay amplitude proportional to $\lambda_u^{(s)} \sim \lambda^4$
is neglected. The operators  ${\cal O}_i$ in (\ref{eq:Heff}) 
and the numerical values of their Wilson coefficients
$C_i$ used in this paper are listed in the Appendix~A of Ref.~\cite{Hambrock:2015wka}
and in Table~\ref{tab:ci} above.
In the decay amplitude (\ref{eq:ampl})
the dominant contributions of the operators ${\cal O}_{9,10}$ 
and ${\cal O}_7$  are factorized to the $B\to P$ form factors. 
The additional amplitudes denoted as  ${\cal H}_{BP}^{(c)} (q^2), 
{\cal H}_{BP}^{(u)} (q^2)$  in Eqs.~(\ref{eq:amplc}), (\ref{eq:amplh}) accumulate 
the  nonlocal effects generated by the all remaining effective operators 
combined with the electromagnetic emission
of the lepton pair. They can be represented as a 
correlation function
of the time-ordered product of effective operators with the quark
e.m. current,  
$j_\mu^{\rm em} = \sum_{q = u,d,s,c,b} Q_q \bar q \gamma_\mu q$,
sandwiched between $B$ and $P$ states:   
\begin{eqnarray}
{\cal H}_{(BP)\mu}^{(p)} & =&  i \int d^4 x e^{i q x} \langle P (p) |{\rm T} \biggl\{ j_\mu^{\rm em} (x), \biggl[C_1 {\cal O}_1^{p} (0) + C_2 {\cal O}_2^{p} (0)
\label{eq:corr} \\
+ & & \hspace*{-7mm}\sum\limits_{k=3-6,8g} C_k {\cal O}_k (0) \biggr] \biggr\} | B (p+q) \rangle = \left[ (p \cdot q) q_\mu - q^2 p_\mu \right] {\cal H}_{BP}^{(p)} (q^2),~~(p=u,c).
\nonumber
\end{eqnarray}
In the case of the $B \to K \ell^+ \ell^-$ decay 
only the amplitude ${\cal H}_{BK}^{(c)} (q^2)$ contributes.
The calculation of the nonlocal amplitudes following the method 
suggested in \cite{Khodjamirian:2010vf} proceeds in two stages. First,
the amplitudes ${\cal H}_{BP}^{(c,u)} (q^2)$  are splitted in the 
contributions with different topologies, including $c$ or $u$ quark 
emission in LO, NLO factorizable corrections, nonfactorizable 
effects of soft gluon emission, hard-spectator 
and annihilation contributions. They are calculated
one by one at spacelike $q^2<0$ where the light-cone OPE for the correlation function  (\ref{eq:corr}) is valid. For the hard-gluon NLO and spectator contributions we apply 
the QCD factorization and for the soft gluon emission the dedicated LCSRs. 
A detailed account of this calculation can be found in   
Refs.~\cite{Khodjamirian:2012rm} and \cite{Hambrock:2015wka}.
After that, the resulting functions ${\cal H}_{BP}^{(c,u)} (q^2<0)$
are fitted to the hadronic dispersion  relations in the $q^2$ variable
where the contributions  
from the lowest vector mesons $V=\rho,\omega,\phi,J/\psi$, $\psi(2S)$
are isolated and the excited states and continuum contributions are modeled,
employing the quark-hadron duality. 
Here we employ as an additional input the experimental  
data on branching fractions of the nonleptonic $B\to VP$ decays  
determining  together with the vector meson decay constants
the  moduli of the residues in the pole terms of the dispersion
relation. 
The phases of these contributions are included in 
the set of fit parameters. Since in this paper 
we are interested only in the large recoil 
(low $q^2$) region, the integral over hadronic spectral density 
at $q^2>4m_D^2$ with no singularities in the large recoil region 
is modeled by a polynomial with complex parameters 
(see Refs.\cite{Khodjamirian:2012rm,Hambrock:2015wka} for details). Indeed, 
for our purposes it is not necessary to use a more 
detailed hadronic representation, like the ansatz suggested 
in \cite{Lyon:2014hpa} and used in \cite{Aaij:2016cbx},
where the broad charmonium resonances located above the open charm 
threshold are resolved with separate relative phases.     

Having fitted the parameters of dispersion relations, we continue 
them to the positive values of $q^2$ in the large recoil region, where 
there is a minor influence of the model-dependent contributions. 
Finally, we note that in our approach 
the differences between the $B_s \to K$, $B \to K$ and 
$B \to \pi$  nonlocal amplitudes originate from the  $SU(3)_{fl}$ -violating 
differences between the decay constants, parameters of light-meson DAs 
and  nonleptonic amplitudes, as well as from the different 
spectator-quark flavours, determining the diagram content of 
these  amplitudes.

\end{document}